\documentclass[11pt,twocolumn,twocolappendix]{aastex631}

\definecolor{gabonavy}{RGB}{0,0,128}

\usepackage{tipa}

\begin{document}

\title{The James Webb Interferometer: Space-based interferometric detections of PDS 70 b and c at 4.8 $\mu$m}
\author[0000-0001-9582-4261]{Dori Blakely}
\affiliation{Department of Physics and Astronomy, University of Victoria, 3800 Finnerty Road, Elliot Building, Victoria, BC, V8P 5C2, Canada}
\affiliation{NRC Herzberg Astronomy and Astrophysics, 5071 West Saanich Road, Victoria, BC, V9E 2E7, Canada}

\author[0000-0002-6773-459X]{Doug Johnstone}
\affiliation{NRC Herzberg Astronomy and Astrophysics,
5071 West Saanich Road,
Victoria, BC, V9E 2E7, Canada}
\affiliation{Department of Physics and Astronomy, University of Victoria, 3800 Finnerty Road, Elliot Building, Victoria, BC, V8P 5C2, Canada}

\author[0000-0001-7255-3251]{Gabriele Cugno}
\affiliation{Department of Astronomy, University of Michigan, Ann Arbor, MI 48109, USA}

\author[0000-0003-1251-4124]{Anand Sivaramakrishnan}
\affiliation{Space Telescope Science Institute, 3700 San Martin Drive, Baltimore, MD 21218, USA}
\affiliation{Astrophysics Department, American Museum of Natural History, 79th Street at Central Park West, New York, NY 10024}
\affiliation{Department of Physics and Astronomy, Johns Hopkins University, 3701 San Martin Drive, Baltimore, MD 21218, USA}

\author[0000-0001-7026-6291]{Peter Tuthill}
\affiliation{Sydney Institute for Astronomy, School of Physics, University of Sydney, NSW~2006, Australia}

\author[0000-0001-9290-7846]{Ruobing Dong}
\affiliation{Department of Physics and Astronomy, University of Victoria, 3800 Finnerty Road, Elliot Building, Victoria, BC, V8P 5C2, Canada}

\author[0000-0003-2595-9114]{Benjamin J. S. Pope}
\affiliation{School of Mathematics and Physics, University of Queensland, St Lucia, QLD~4072, Australia}
\affiliation{University of Southern Queensland, Centre for Astrophysics, Toowoomba, Queensland, Australia}

\author[0000-0003-0475-9375]{Lo\"ic Albert}
\affiliation{Trottier Institute for Research on Exoplanets, Département de Physique, Université de Montréal, 1375 Ave Thérèse-Lavoie-Roux, Montréal, QC, H2V 0B3, Canada}

\author[0009-0003-5950-4828]{Max Charles}
\affiliation{Sydney Institute for Astronomy, School of Physics, University of Sydney, NSW~2006, Australia}

\author[0000-0001-7864-308X]{Rachel A. Cooper}
\affiliation{Space Telescope Science Institute, 3700 San Martin Drive, Baltimore, MD 21218, USA}

\author[0000-0003-1863-4960]{Matthew De Furio}
\affiliation{Department of Astronomy, The University of Texas at Austin, 2515 Speedway Stop C1400, Austin, TX 78712, USA}

\author[0000-0002-1015-9029]{Louis Desdoigts}
\affiliation{Sydney Institute for Astronomy, School of Physics, University of Sydney, NSW~2006, Australia}

\author[0000-0001-5485-4675]{Ren\'e Doyon} 
\affiliation{Trottier Institute for Research on Exoplanets, Département de Physique, Université de Montréal, 1375 Ave Thérèse-Lavoie-Roux, Montréal, QC, H2V 0B3, Canada}
\affiliation{Observatoire du Mont-M\'egantic, Universit\'e de Montr\'eal, Montr\'eal H3C 3J7, Canada}

\author[0000-0001-8822-6327]{Logan Francis}
\affiliation{Leiden Observatory, Leiden University, PO Box 9513, 2300 RA Leiden, The Netherlands}

\author[0000-0002-7162-8036]{Alexandra Z. Greenbaum}
\affiliation{IPAC, Caltech, 1200 E. California Blvd., Pasadena, CA 91125, USA}

\author[0000-0002-6780-4252]{David Lafreni\`ere}
\affiliation{Trottier Institute for Research on Exoplanets, Département de Physique, Université de Montréal, 1375 Ave Thérèse-Lavoie-Roux, Montréal, QC, H2V 0B3, Canada}

\author[0009-0009-9434-8860]{James P. Lloyd}
\affiliation{Carl Sagan Institute/Department of Astronomy, Cornell University, Ithaca NY 14853}

\author[0000-0003-1227-3084]{Michael R. Meyer}
\affiliation{Department of Astronomy, University of Michigan, Ann Arbor, MI 48109, USA}

\author[0000-0003-3818-408X]{Laurent Pueyo}
\affiliation{Space Telescope Science Institute, 3700 San Martin Drive, Baltimore, MD 21218, USA}

\author[0000-0003-2259-3911]{Shrishmoy Ray}
\affiliation{School of Mathematics and Physics, University of Queensland, St Lucia, QLD~4072, Australia} 

\author[0000-0002-9723-0421]{Joel S\'anchez-Berm\'udez}
\affiliation{Instituto de Astronom\'ia, Universidad Nacional Aut\'onoma de M\'exico, Apdo. Postal 70264, Ciudad de M\'exico, 04510, M\'exico}
\affiliation{Max-Planck-Institut f\"ur Astronomie, K\"{o}nigstuhl 17, D-69117 Heidelberg, Germany}

\author[0000-0001-7661-5130]{Anthony Soulain}
\affiliation{Univ.\ Grenoble Alpes, CNRS, IPAG, 38000 Grenoble, France}

\author[0000-0002-1536-7193]{Deepashri Thatte}
\affiliation{Space Telescope Science Institute, 3700 San Martin Drive, Baltimore, MD 21218, USA}

\author[0000-0001-5684-4593]{William Thompson}
\affiliation{NRC Herzberg Astronomy and Astrophysics,
5071 West Saanich Road,
Victoria, BC, V9E 2E7, Canada}

\author[0000-0002-5922-8267]{Thomas Vandal}
\affiliation{Trottier Institute for Research on Exoplanets, Département de Physique, Université de Montréal, 1375 Ave Thérèse-Lavoie-Roux, Montréal, QC, H2V 0B3, Canada}



\begin{abstract}

We observed the planet-hosting system PDS 70 with the James Webb Interferometer, JWST's Aperture Masking Interferometric (AMI) mode within NIRISS. Observing with the F480M filter centered at 4.8 $\mu$m, we simultaneously fit {geometrical models} 
to the outer disk and the two known planetary companions. We re-detect the protoplanets PDS 70 b and c at an {SNR of 14.7 and 7.0}, respectively. Our photometry of both PDS 70 b and c {provides tentative evidence of mid-IR} circumplanetary disk emission through fitting SED models to these new measurements and those found in the literature. We also newly detect emission within the disk gap at an SNR of $\sim$4, {at a position angle of $220^{+10}_{-15}$ degrees}, 
and an unconstrained separation within $\sim$200 mas. Follow-up observations will be needed to determine the nature of this emission. {We place a 5$\sigma$ upper limit of 208 $\pm$ 10 $\mu$Jy 
on the flux }
of the candidate PDS 70 d at 4.8 $\mu$m, which indicates that if the previously observed emission at shorter wavelengths is due to a planet, this putative planet has a different atmospheric composition than PDS 70 b or c. Finally, we place upper limits on emission from any additional planets in the disk gap. {We find an azimuthally averaged 5$\sigma$ contrast upper limit $>$7 magnitudes at separations greater than 110 mas}. These are the deepest limits to date within $\sim$250 mas at 4.8 $\mu$m and the first space-based interferometric observations of this system. 


\end{abstract}



\section{Introduction} \label{sec:intro}

PDS 70 is one of the most extensively studied young stellar systems. It is the only known multi-planet protoplanetary disk system, where two or more planets have been robustly detected within the disks from which they formed \citep{kep1,2019NatAs...3..749H}. 
The disk consists of a large outer component and a smaller inner component \citep{kep1}, separated by a wide cavity \citep{2012ApJ...758L..19H,2021AJ....162...99F} spanning between $\lesssim$17\,au and $\sim$54\,au \citep{kep1}.
The inner disk was first inferred via spectral energy distribution (SED) fitting \citep{2012ApJ...758L..19H,2012ApJ...760..111D}, and was subsequently isolated in the near-infrared \citep{kep1} and sub-millimeter \citep{kep2}. Recently observations with JWST/MIRI detected water vapour in the inner disk \citep{2023Natur.620..516P}. The outer disk of PDS 70 is not significantly perturbed in contrast to some other protoplanetary disk systems where planet candidates have been detected, such as MWC 758 \citep{2023NatAs...7.1208W} or AB Aurigae \citep{2022NatAs...6..751C}. The PDS 70 outer disk appears as a close to symmetric ring at sub-mm wavelengths \citep{kep2,2021ApJ...916L...2B}, with one significant azimuthal asymmetry observed to the {north-west} of the star that is close to coincident with an arm-like structure seen in the near-infrared \citep{2020AJ....159..263W,2022A&A...668A.125J}.

The planets PDS 70 b and c have been extensively imaged in the near-infrared from $\sim$1 to 3.8 $\mu$m \citep[e.g.,][]{kep1,2018A&A...617L...2M,2019A&A...632A..25M,2019ApJ...877L..33C,2020AJ....159..263W,2021AJ....161..148W,2021A&A...653A..12C}, with a few detections beyond $\sim$4 $\mu$m to 4.8 $\mu$m \citep{2020A&A...644A..13S,2024arXiv240304855C}, as well as detections in H$_{\alpha}$ \citep{2018ApJ...863L...8W,2019NatAs...3..749H}. Circumplanetary disk emission from PDS 70 c has been directly detected in the sub-mm with ALMA, coincident with detections in the near-infrared \citep{2019ApJ...879L..25I,2021ApJ...916L...2B}. Sub-mm emission has also been observed that is seemingly associated with PDS 70 b, however it is offset from the near-infrared detections \citep{2019ApJ...879L..25I,2021ApJ...916L...2B}. Additionally, emission at the L$_5$ Lagrangian point of PDS 70 b has tentatively been detected in the sub-mm with ALMA \citep{2023A&A...675A.172B}. Finally, a third point source has been tentatively detected in the PDS 70 disk cavity, using VLT/SPHERE over multiple epochs \citep{2019A&A...632A..25M}. The nature of the source is still not well understood as it has a distinct spectrum compared to both PDS 70 b and c, more similar to the spectrum of PDS 70 A between $\sim$1-1.6 $\mu$m, implying it may be scattered light \citep{2019A&A...632A..25M}. This source was tentatively re-detected at 1.9 $\mu$m with JWST/NIRCam \citep{2024arXiv240304855C}.







{Beyond $\sim$4 $\mu$m, emission from warm ($\gtrsim$100 K) circumplanetary disk (CPD) material is expected to contribute comparably (and greater at longer wavelengths) to atmospheric emission \citep[e.g.,][]{2021AJ....161..148W,2024arXiv240407086C}. So, observations into the mid infrared may allow for the presence of circumplanetary disk emission to be inferred by looking for excess emission compared to atmospheric models.}
\cite{2021AJ....161..148W} showed that the VLT/NACO M$'$ detection of PDS 70 b from \cite{2020A&A...644A..13S} provided weak support for excess blackbody-like emission, so follow up observations at $\sim$4.8 $\mu$m and beyond will allow for this tentative detection of CPD emission in the mid-infrared to be confirmed. Recently, \cite{2024arXiv240304855C} re-detected PDS 70 b and c with the JWST/NIRCam (4.8 $\mu$m) F480M filter. Their detection of PDS 70 b was in agreement with the M$'$ detection, but with a larger uncertainty. 

In our work, we re-detect PDS 70 b and c at F480M using {the Aperture Masking Interferometry mode on JWST/NIRISS}, 
an independent method to {the JWST/NIRCam results, presented by \cite{2024arXiv240304855C}}, and make the most precise measurement of the flux of PDS 70 b at 4.8 $\mu$m, using the JWST/NIRISS F480M filter. We also detect PDS 70 c using the F480M filter at a similar precision to \cite{2024arXiv240304855C}. 
{Here, using the power of {NIRISS/AMI}, as introduced in these previous works using AMI: \citet{2024ApJ...963..127L,2023arXiv231011508R,2023arXiv231011499S}, 
we detect PDS 70, its outer disk, and its two protoplanets, b and c. These are the first planets detected with space-based interferometry.}

This paper is structured as follows: \S\,\ref{sec:obs} outlines the data acquisition, reduction and cleaning procedure; \S\,\ref{sec:met} details the disk plus two planet model fitting and the conversion of the measured contrasts to fluxes; \S\,\ref{sec:res} outlines the derived planet parameters and the analysis of the data for any signal beyond the two known planets and the disk; \S\,\ref{sec:disc} discusses the implications of the measured contrasts and the nature of the tentative detection of residual emission in the data. We close with a summary in \S\,\ref{sec:con}.

\section{Observations and data reduction} \label{sec:obs}

We observed PDS 70 as a part of the NIRISS GTO program (PID 1242, PI D. Johnstone). The data presented in this paper were acquired on February 24 2023 using the James Webb Interferometer, 
{the aperture masking interferometry mode of the NIRISS instrument \citep{2012SPIE.8442E..2SS,2023PASP..135a5003S,2023PASP..135i8001D} henceforth referred to as AMI, in the F480M filter ($\lambda=4.815$ $\mu$m, $\Delta\lambda=0.298$ $\mu$m)}\footnote{https://jwst-docs.stsci.edu/jwst-near-infrared-imager-and-slitless-spectrograph/niriss-instrumentation/niriss-filters}.
The SUB80 sub-array was used along with the NISRAPID readout pattern. The point-spread-function (PSF) calibrator star HD 123991 was also observed using the same configuration. Both targets were observed using no dithers and no rolls. The PDS 70 data set consists of 96 groups and 418 integrations, while the brighter HD 123991 data set consists of 29 groups and 418 integrations. This corresponds to exposure times of 50.45 minutes for PDS 70 and 15.24 minutes for HD 123991.

\begin{figure}
\centering
\includegraphics[width=1\linewidth]{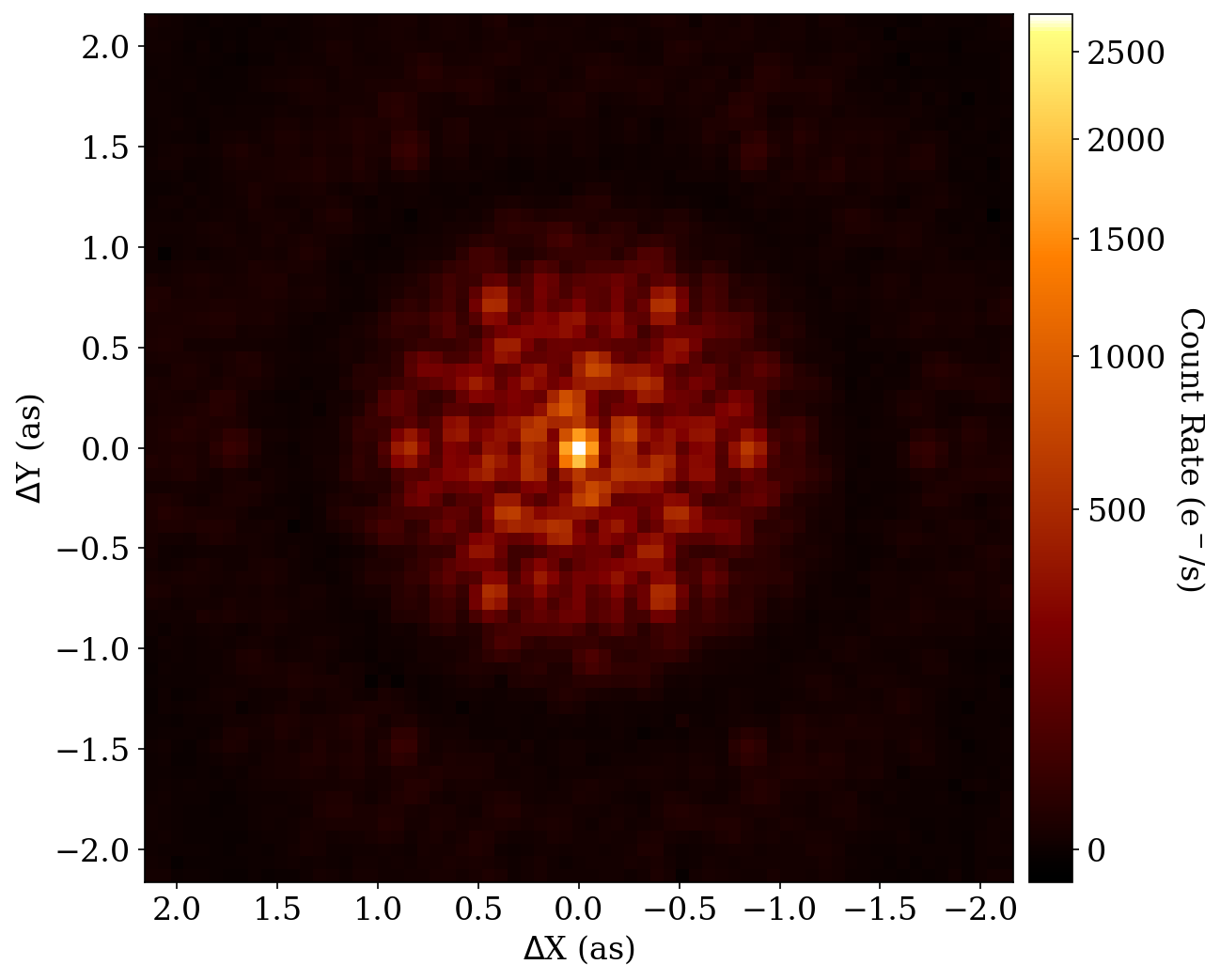}
\caption{Mean PDS 70 \texttt{calints} file, after bad pixel correction, using only the first 28 groups. }
\label{fig:f480M_data}
\end{figure}

The data were reduced using version 1.11.3 of the \texttt{jwst} pipeline\footnote{https://github.com/spacetelescope/jwst} \citep{bushouse_2023_8157276}, {with Calibration Reference Data System context \texttt{jwst\_1110.pmap}}, from the \texttt{uncal} data format. We use the default \texttt{calwebb\_detector1} stage 1 pipeline and \texttt{calwebb\_image2} stage 2 pipeline, {skipping only the \texttt{IPC}, \texttt{photom} and \texttt{resample} steps}.
Note that we do not use the \texttt{charge\_migration}\footnote{named \texttt{undersampling\_correction} in the 1.11.3 version of the pipeline that we used} step \citep{2024PASP..136a4503G}, that was designed to minimize the impact of the detector brighter-fatter effect (BFE). The \texttt{charge\_migration} algorithm discards all data in each pixel above a set signal threshold, along with the rest of the data in the four directly adjacent neighbors. This strategy is not optimal for AMI data because, as we show in Appendix \ref{sec:bfe}, and has been shown in other works \citep[e.g.,][]{2023A&A...680A..96A}, charge bleeding from a bright pixel, into its neighbors, affects not only the four adjacent pixels, but also the four pixels along the diagonals, albeit somewhat less. Additionally, considering only the brightness of individual pixels is not the optimal threshold criteria because the BFE depends on the contrasts between pixels \citep[e.g.,][]{2017JInst..12C4009P}. Due to the fine structure in the AMI PSF, that is somewhat undersampled at 4.8 $\mu$m with the NIRISS plate scale of $\sim$66 mas, there are many parts of the PSF with large contrasts between neighboring pixels. The current pipeline BFE mitigation algorithm, therefore, is not useful for correcting AMI data. A modification to the existing pipeline, beyond the scope of this work, could be to base the threshold on the contrast between adjacent and diagonal neighboring pixels.

We investigate the effect of the BFE on the data (see Appendix \ref{sec:bfe} for details), as it has been shown to be the limiting noise floor for AMI observations \citep{2023PASP..135a5003S,2023PASP..135a4502K,2023arXiv231011508R,2023arXiv231011499S}. Based on our analysis, and to minimize the effect of the BFE on the calculated squared visibilities and closure phases, we discard all groups beyond where the sum of the central 3$\times$3 pixels of the PSF is above 30,000 data number (DN), or $\sim$48,000 electrons, in the linearized ramp-level data. We use this cutoff, as opposed to the intensity in the central pixel, because it is more robust against changes in PSF centering. This constraint corresponds to an apparent systematic change in the calculated rate in the central pixel of $\sim$1\%, and minimal bleeding into surrounding pixels (Appendix \ref{sec:bfe}, Figure \ref{fig:pix_ev}). We note that this loss in signal biases our measured contrasts by significantly less than 1\%, due to the central pixel accounting for $\sim$1\% of the total signal in the PSF. {This can be seen in Figure \ref{fig:f480M_data}, which presents the mean PDS 70 image-plane data using only the first 28 groups.} We also note that this BFE mitigation method is only possible because PDS 70 and HD 123991 are relatively faint, such that we have at least 10 groups before a peak signal level of $\sim$5800 DN. With fewer groups, the data is likely to be dominated by the 1/f noise seen in the NIRISS detector \citep[e.g.,][]{2023arXiv231011499S}. For PDS 70 we are therefore limited to 28 out of the total 96 groups, while for HD 123991 we are limited to 10 out of the total 29 groups.

We correct bad-pixels that were flagged by the pipeline by finding the pixel value that minimizes the power outside of the Fourier support of the AMI mask \citep{2013MNRAS.433.1718I}. We also perform sub-pixel centering of the data using a Fourier-based shift, by finding the position that minimizes the absolute value of the phase calculated from the mean of the cleaned data \citep{2019MNRAS.486..639K}.

The background star near PDS 70 that was originally observed by \cite{2006A&A...458..317R}, and confirmed to be a background star by \cite{2012ApJ...758L..19H}, is at the edge of the SUB80 array, at a separation of {$\sim$2 arc seconds (as) from PDS 70 A}. This is outside of the interferometric field-of-view, set by the shortest baseline, but the signal will still contaminate all baselines. To minimize the contribution to the calculated interferometric observables, a super-Gaussian window ($e^{-(r/\sigma)^4}$), with $\sigma$ = 30 pixels,  
is applied to both the PDS 70 data and the HD 123991 data. 

\begin{figure*}
\centering
\includegraphics[width=0.85\linewidth]{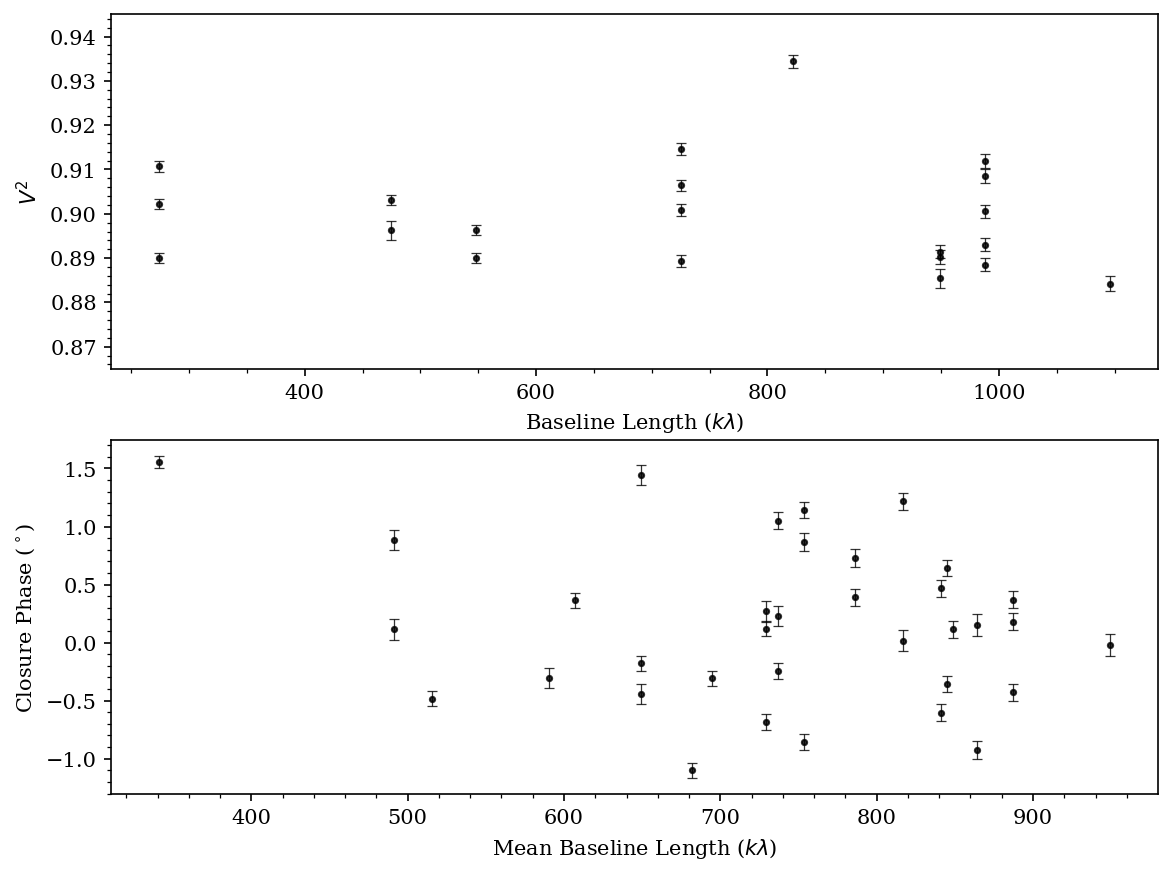}
\caption{Squared visibility data (top) and the closure phase data (bottom).}
\label{fig:f480M_cps}
\end{figure*}

We calculate the squared visibilities, $V^2$, and closure phases, of PDS 70 and calibrator, HD 123991, using \texttt{amical} \citep{2020SPIE11446E..11S,2023ascl.soft02021S}. We also compute the statistical uncertainties (i.e. the standard error of the mean) of these quantities across integrations. The interferometric observables are calibrated, to remove the instrumental bias, by dividing the $V^2$ of PDS 70 by the $V^2$ of HD 123991. Similarly, the closure phases are calibrated by subtracting the closure phases of HD 123991 from those of PDS 70. The calibrated closure phases and squared visibilities are shown in Figure \ref{fig:f480M_cps}.

\section{Methods} 
\label{sec:met}

Due to the interferometric nature of the data, we construct a model to jointly fit for flux from the star, extended disk emission, and the two known planets. This is required because the signal from all components is entangled and thus cannot be measured independently. We model the star and the two planets as delta functions, with their analytic Fourier transform given by
\begin{equation}
    V_{*,b,c}(u,v) = I_* + I_b e^{-i 2 \pi (ux_b + vy_b)} + I_c e^{-i 2 \pi (ux_c + vy_c)},
    \label{eqn:pts_vis_eqn}
\end{equation}
where $I_*$ is the brightness of the star, $I_b$ is the brightness of planet b, and $I_c$ is the brightness of planet c. Additionally, $x_b$, $y_b$, $x_c$ and $y_c$ denote the $x$ and $y$ offset of planets b and c relative to the star (which is fixed at the phase center), and $u$ and $v$ are the baseline coordinates in the Fourier domain, in units of wavelength.

To model the extended disk emission, we use a simple geometrical model, as {has been used to successfully} reconstruct the extended emission of LkCa 15, in the near infrared, with ground based AMI \citep{2022ApJ...931....3B,2023ApJ...953...55S}. A simple model is sufficient due to the sparsity of the data, and the fact that the dominant signal is expected to be from the forward scattering peak of the disk. To construct this model, we assume that all of the disk signal that we are observing is {either scattered light or} optically thick emission from the disk surface, in a similar manner to the procedure used by \cite{2016A&A...596A..70S} and \cite{2018A&A...619A.160S}. We describe the height of the surface of the disk using a power law profile, given by
\begin{equation}
\label{eqn:z}
    z = H_{100}\left(\frac{\sqrt{x^2 + y^2}}{100 \mathrm{\: au}}\right)^{\beta},
\end{equation}
where x and y are the potentially rotated and/or inclined coordinates of the disk mid-plane, in units of au, $H_{100}$ is the aspect ratio at 100 au ($\sim$0.9 arc seconds), and $\beta$ is the flaring angle. To project this geometry into the plane of the sky, we define $r$ to be the apparent distance from the star to the disk surface, given an observed inclination angle, $i$, relative to the plane of the sky,

\begin{equation}
\label{eqn:rad}
    r = \sqrt{x^2 + (y + z \sin(i))^2 + z^2}.
\end{equation}

\noindent We describe the radial brightness distribution as a skewed Gaussian projected onto the flared disk geometry, given by $I_{d_1}(r) \cdot I_{d_2}(r)$, where $I_{d_1}(r)$ is defined as

\begin{equation}
\label{eqn:odradial}
    I_{d_1}(r) = \exp\left(-\frac{(r-r_0)^2}{2\sigma_r^2}\right),
\end{equation}
and $I_{d_2}(r)$ is
\begin{equation}
\label{eqn:odskew}
    I_{d_2}(r) = \frac{1}{2} \left(1 + \mathrm{erf}\left(\alpha \frac{(r-r_0)}{\sqrt{2}\sigma_r} \right)\right),
\end{equation}
where $r_0$ describes the location of the peak of the disk brightness, $\sigma_r$ describes the radial extent of the disk emission and $\alpha$ describes the degree to which the inner edge of the disk is truncated.

The azimuthal brightness distribution is {the sum of an axisymmetric component and an asymmetric component. For the asymmetric component, we explore three independent parametrizations: a Henyey-Greenstein model \citep{1941ApJ....93...70H}, a Gaussian model, and a power law model. The Henyey-Greenstein model is given by
\begin{equation}
\label{eqn:odazimuthal}
    I_{d_3}(\theta) = \frac{1-g^2}{4 \pi (1 + g^2 -2 g \cos(\theta - \theta_0))^{3/2}},
\end{equation}
where $g$ is the scattering parameter, a free parameter constrained to be between 0 (isotropic scattering) and 1 (forward scattering), $\theta$ is the azimuthal angle in the disk mid-plane (defined by the coordinates x and y), and $\theta_0$ is the azimuthal location of the scattering peak, which is fixed to the minor axis on the near side of the disk. The Gaussian model is given by 
\begin{equation}
\label{eqn:odazimuthal_Gaussian}
    I_{d_3}(\theta) = \exp\left(-\frac{(\theta-\theta_0)^2}{2\sigma_{\theta}^2}\right),
\end{equation}
where $\sigma_{\theta}$ describes the azimuthal extent of the disk emission. The power law model is given by
\begin{equation}
\label{eqn:odazimuthal_powerlaw}
    I_{d_3}(\theta) = \cos\left( \frac{1}{2}(\theta-\theta_0)\right)^N,
\end{equation}
where $N$ describes the azimuthal extent of the disk emission.}


Combining Equations \ref{eqn:z}
through \ref{eqn:odskew}, and any one of scattering phase functions given by Equations \ref{eqn:odazimuthal}, \ref{eqn:odazimuthal_Gaussian}, or \ref{eqn:odazimuthal_powerlaw}, the full disk model is given by
\begin{equation}
\label{eqn:model}
    I_{d}(r,\theta) = \left(A_a I_{d_3}(\theta) + A_s \right)  I_{d_1}(r) I_{d_2}(r),
\end{equation}
where $A_a$ and $A_s$ control the brightness of the asymmetric and symmetric components, respectively. The Fourier transform of Equation \ref{eqn:model} is calculated using the one-sided discrete Fourier transform (DFT) as implemented in \texttt{XARA} \citep{2010ApJ...724..464M,2013PASP..125..422M,2020A&A...636A..72M}, to exactly compute the DFT at the baseline coordinates of the NIRISS aperture mask baselines.  

Finally, any over-resolved emission is included in the model using a single free parameter $I_o$, that contributes additional flux to the normalization of the visibilities. The full model consists of the sum of Equation \ref{eqn:pts_vis_eqn} and the DFT of Equation \ref{eqn:model}, normalized by the total flux of the model. The complex visibilities of {these 17 parameter models} are given by
\begin{equation}
\label{eqn:modelvis}
    V_{m} = \frac{\mathcal{F}\{I_{d}\} + V_{*,b,c}}{\Sigma I_{d} + I_* + I_b + I_c + I_o}.
\end{equation}
We calculate the closure phase of the model by summing the phase measured between baselines that form a (closure) triangle, for all 35 triplets of holes in the aperture mask, corresponding to 15 independent measurements. Squared visibilities are the squared amplitude of the complex visibilities, corresponding to each of the 21 baselines formed by pairs of mask holes. 

{To fit our model to the data, we further construct a more robust set of observables from the squared visibilities and closure phases, following a similar approach to the Bayesian analysis in \cite{Xuan2024} (and described in more detail in \cite{GL229B_methods}). We calculate self-calibrating log-closure amplitudes, which calibrate out hole-based gain terms affecting the squared visibilities. We calculate the log-closure amplitude uncertainties using a Monte Carlo approach, by calculating the standard deviation of 1,000,000 noise realizations from the statistical uncertainties of the squared visibilities. 
We also consider correlations between shared baselines, using analytic correlation matrices for the log-closure amplitudes and closure phases. For the closure phases, we use the closure phase correlation matrix, $C_{\Psi}$, described by \cite{2020A&A...644A.110K} (with no spectral component). We construct a similar log-closure amplitude correlation matrix, $C_{a}$, following \cite{2020ApJ...894...31B}. The closure phase and log-closure amplitude correlation matrices can be constructed using Equations B5 and B14 derived by \cite{2020ApJ...894...31B}, respectively, setting $\sigma = 1/\sqrt{3}$ for closure phases and $\sigma = 1/\sqrt{4}$ for log-closure amplitudes. Finally, we construct linearly independent closure phase and log-closure amplitude observables. We calculate projection matrices by taking the Singular Value Decomposition (SVD) of $T T^T$, where $T$ is the design matrix of either the closure phases or log-closure amplitudes, as defined by \cite{2020ApJ...894...31B}, and only taking the non-zero singular values. This is similar to what is described by  \cite{2013MNRAS.433.1718I} for closure phases. We henceforth refer to these new observables as kernel phases and log kernel amplitudes. The advantage of using the ``kernel" observables is twofold. By using closure phases, we are ``double counting" the data in the likelihood calculation, leading to overly confident results and underestimated parameter uncertainties. The kernel approach also allows for the covariance matrices, described above, to be used in the likelihood, as the matrices are rank-deficient, whereas this is not the case in the new basis.}

{We use a Bayesian modelling approach to measure the locations and contrasts of PDS 70 b and c, as well as our disk model parameters. We estimate our model posterior using dynamic nested sampling \citep{dyn2018}, with \texttt{dynesty} \citep{2020MNRAS.493.3132S,sergey_koposov_2023_7600689}.} {We use uniform priors on the contrasts, separations and position angles of both planets. Our contrast prior ranges from 0 to 10 magnitudes. Our location priors for PDS 70 b are from 100-250 mas in separation and 70-160 degrees in position angle. Similarly, for the position of PDS 70 c, we use a range of 100-250 mas in separation and 230-320 degrees in position angle.} 
We use priors on the the disk geometry from \cite{kep1} and \cite{kep2}, that are described in Appendix \ref{sec:desc_of_geo}. {To account for an unknown level of miscalibration seen in the squared visibilities, which is also present in the log-closure amplitudes, we include a systematic uncertainty, $\sigma_{a,sys}$, term in our model, added in quadrature with the statistical uncertainty, that is recovered along with our star plus disk plus planet model parameters. }

{We estimate parameters for both PDS 70 b and c using Bayesian model averaging over the three asymmetric scattering models we explore.  We follow the approach used by \cite{2024ApJ...966..156N} and \cite{2024A&A...687A.298N}, of assuming that each model is equal likely apriori and thus combine the posterior distributions by weighting each model by their Bayesian evidences (listed in Table \ref{tab:disk_par} in Appendix \ref{sec:desc_of_geo}).}

\begin{figure*}
\centering
\includegraphics[width=0.85\linewidth]{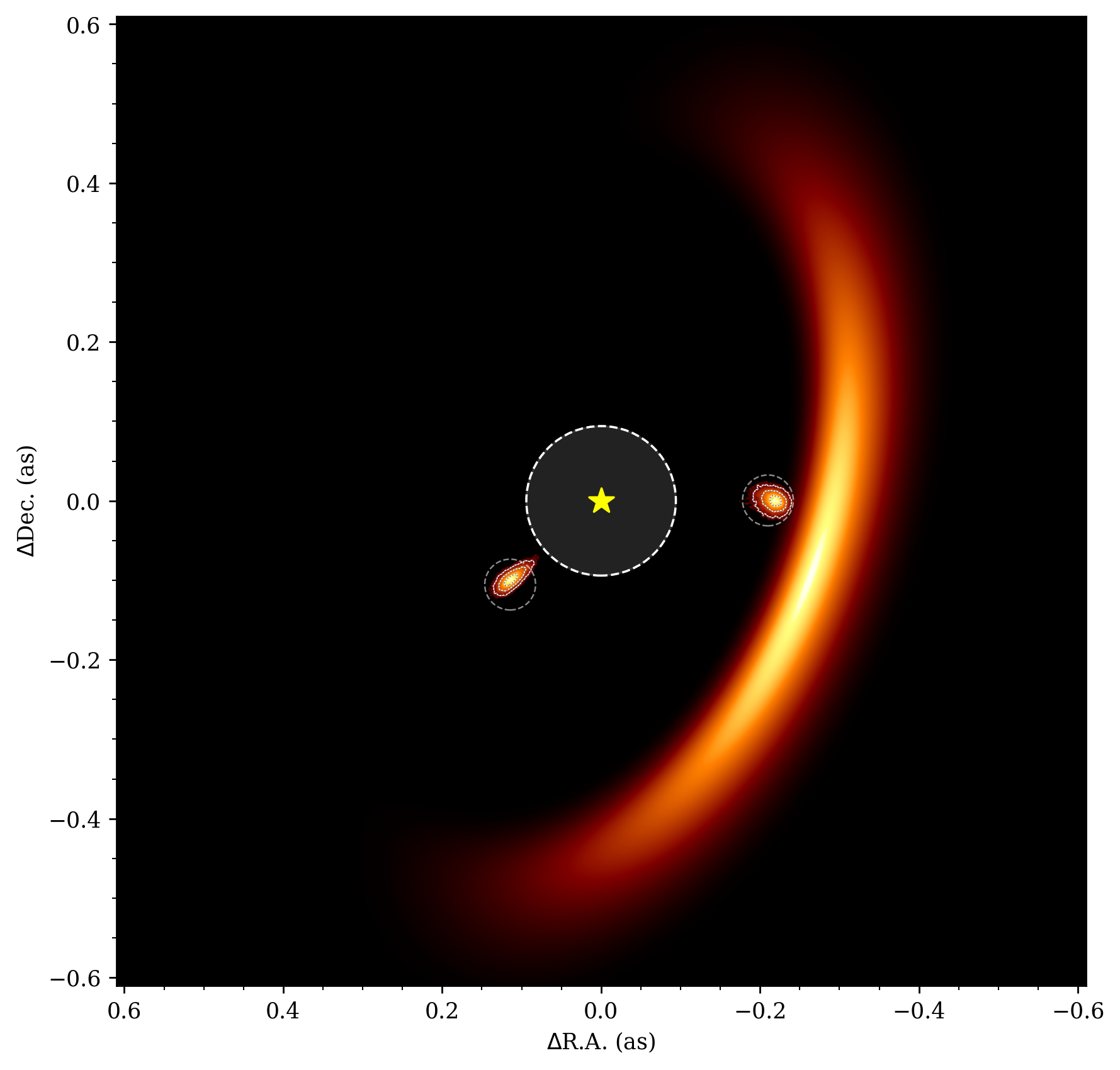}
\caption{{Power law} geometrical model fit to the F480M {kernel phase and log kernel amplitude}
data of PDS 70 (images), using uniform priors on the positions of PDS 70 b and c,{ shown with a linear stretch in arbitrary units}. {With respect to the central source (yellow star), PDS 70 b is to the south east while PDS 70 c is directly west.}
The {white and grey dashed contours, plotted on top of the arbitrarily coloured posterior density,} denote the 1, 2, and 3 $\sigma$ contours of the marginalized posterior of the positions of PDS 70 b and c {from the Bayesian average of the posteriors of the three models}. The two grey dashed circles are centered on the predicted locations of 
the planets
at the time of the observations \citep{2021AJ....161..148W} from \texttt{whereistheplanet} \citep{2021ascl.soft01003W}. We mask the central region to denote the inner working angle of $\sim$0.5$\lambda/B =$ 94\,mas, the diffraction limit of the data.}
\label{fig:f480M_fit}
\end{figure*}

{To convert the measured contrasts to flux measurements, we use aperture photometry. We follow the procedure outlined in Section \ref{sec:obs}, of reducing and cleaning the data. The only difference being that we now apply the \texttt{photom} step, so that the data is in units of surface brightness. We crop the data to a size of 67$\times$67 pixels, with PDS 70 A at the center. We calculate the total flux within a 67 pixel diameter circular aperture and recover a total flux of 128.6 mJy. We note that this value includes resolved disk flux, which can be clearly seen by looking at the squared visibilities decrement from 1, seen at all baselines, in Figure \ref{fig:f480M_cps}. From the visibility amplitudes, we calculate a resolved flux component of 5.2 $\pm$ 0.6 \%, independent of baseline. Additionally, our photometry is missing a non-negligible amount of flux that is outside of the field of view. We calculate a correction factor by calculating PSF models using \texttt{WebbPSF} \citep{2012SPIE.8442E..3DP}. We calculate a 67$\times$67 pixel model and 2000$\times$2000 pixel model, both with the native NIRISS pixel size. We calculate the correction factor by dividing the total flux within a 2000 pixel diameter circular aperture for the full PSF and a 67 pixel diameter for the cutoff PSF, giving a correction factor of 1.12. After applying these correction factors, we find a flux of PDS 70 A (and the unresolved inner disk emission) of 136.4 $\pm$ 6.8 mJy, assuming a 5\% absolute flux calibration uncertainty\footnote{https://jwst-docs.stsci.edu/jwst-calibration-status/niriss-calibration-status/niriss-imaging-calibration-status}.}

\section{Results} 
\label{sec:res}

\subsection{Derived Planet Parameters}

\begin{figure*}
\centering
\includegraphics[width=0.85\linewidth]{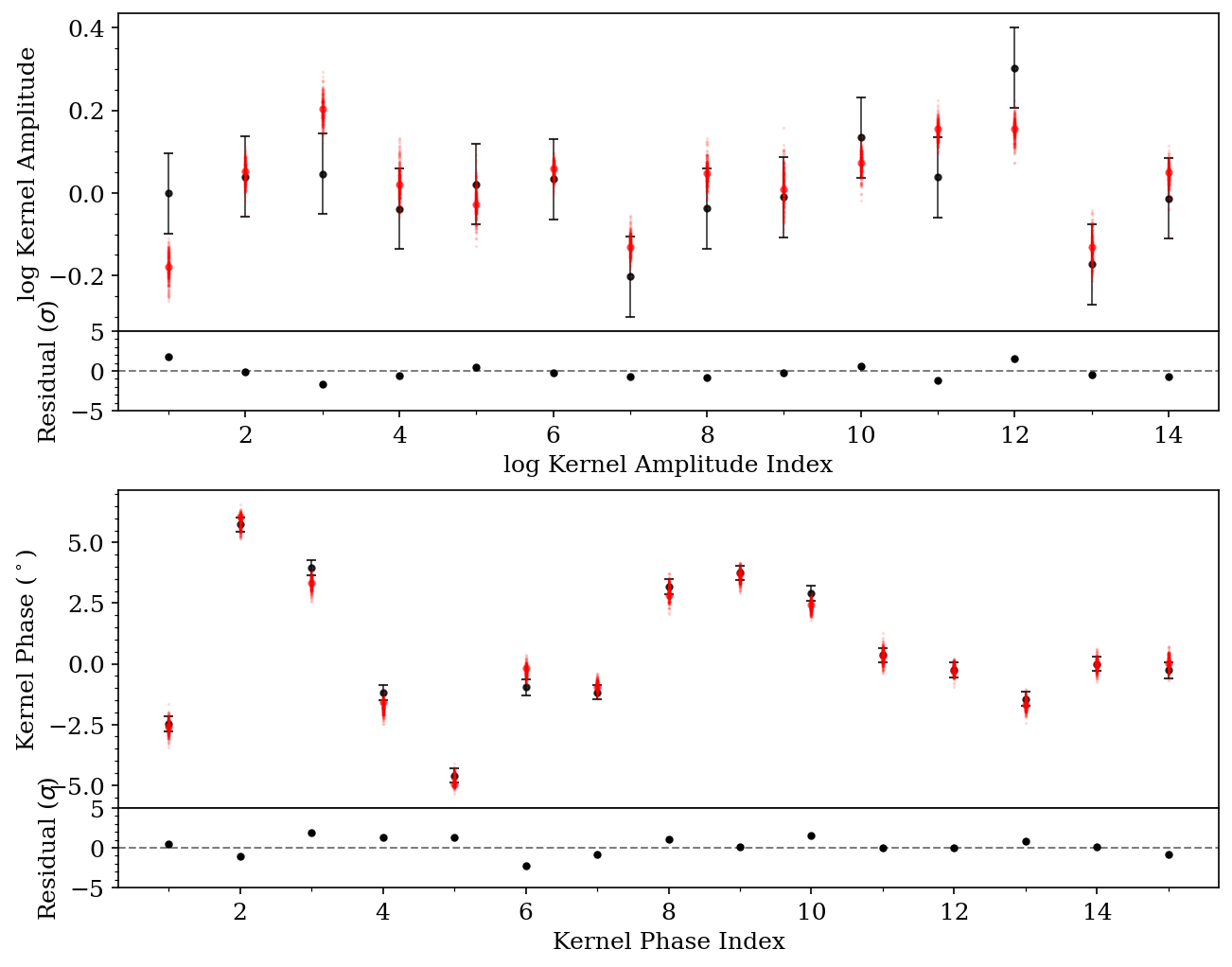}
\caption{
{Log kernel amplitude data, black with the model values plotted in red (top), and the kernel phase data, black, with the model kernel phases shown in red (bottom)}
for the two point source plus {power law} disk geometrical model fit to the F480M data. The error bars are the square root of the diagonal of the covariance matrices, including the estimated systematic uncertainty in the log kernel amplitudes. For both panels, the large red circles denote the values calculated with the median parameter model, and 250 random posterior samples are denoted by the small, transparent red points. The noise normalized residuals are shown below both panels.}
\label{fig:f480Mkernels}
\end{figure*}

Figure \ref{fig:f480M_fit} shows the two planet plus disk  model generated from the median parameters from the {power law scattering model (Equation \ref{eqn:odazimuthal_powerlaw})} model posterior. {We focus on the power law model in particular, because it had the highest Bayesian evidence out of the three models we explored (Table \ref{tab:disk_par} in the Appendix)}. {The 1, 2 and 3 $\sigma$ contours from the marginalized posterior of the planet locations are shown by the white contours. A comparison of the model log kernel amplitudes and kernel phases} 
to the data is given in Figure \ref{fig:f480Mkernels}, showing that the model provides an excellent fit. 
Furthermore, the measured positions of both PDS 70 b and c are consistent with VLTI/GRAVITY predictions from \cite{2021AJ....161..148W}, which are shown by the dashed grey circles in Figure \ref{fig:f480M_fit}.

Table \ref{tab:comp_fit_par_wsys} summarizes the derived planet parameters, calculated from the Bayesian average of all three models. The disk model parameters are listed in Table \ref{tab:disk_par} in Appendix \ref{sec:desc_of_geo}. {We find that both planets are found at a high signal-to-noise ratio (SNR), 14.7 for PDS 70 b and 7.0 for PDS 70 c (where we define the SNR as the contrast divided by the uncertainty in the contrast marginalized over all of the model parameters), in locations that are consistent between the three models (planet parameters for the individual models are shown in Table \ref{tab:comp_fit_par} in the Appendix). We also find that the measured contrasts of the planets are consistent between the three models.} 

\begin{figure*}
\centering
\includegraphics[width=0.9\linewidth]{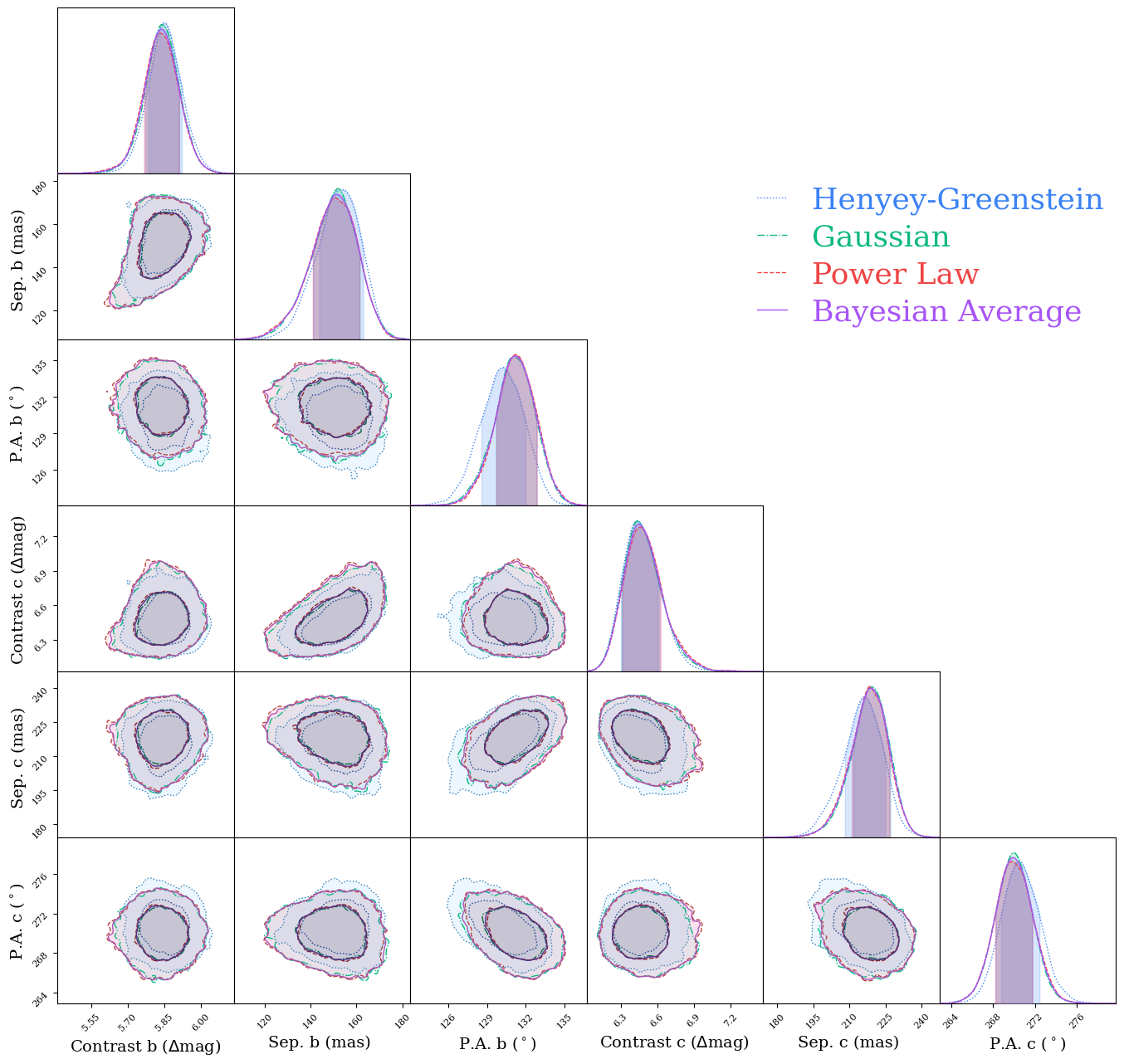}
\caption{Corner plot showing the 1D and 2D marginalized posteriors of the companion parameters from the joint two point source plus geometrical model {fits to the F480M log kernel amplitudes and kernel phases.}
 }
\label{fig:f480M_corner}
\end{figure*}

\begin{figure*}
\centering
\includegraphics[width=1\linewidth]{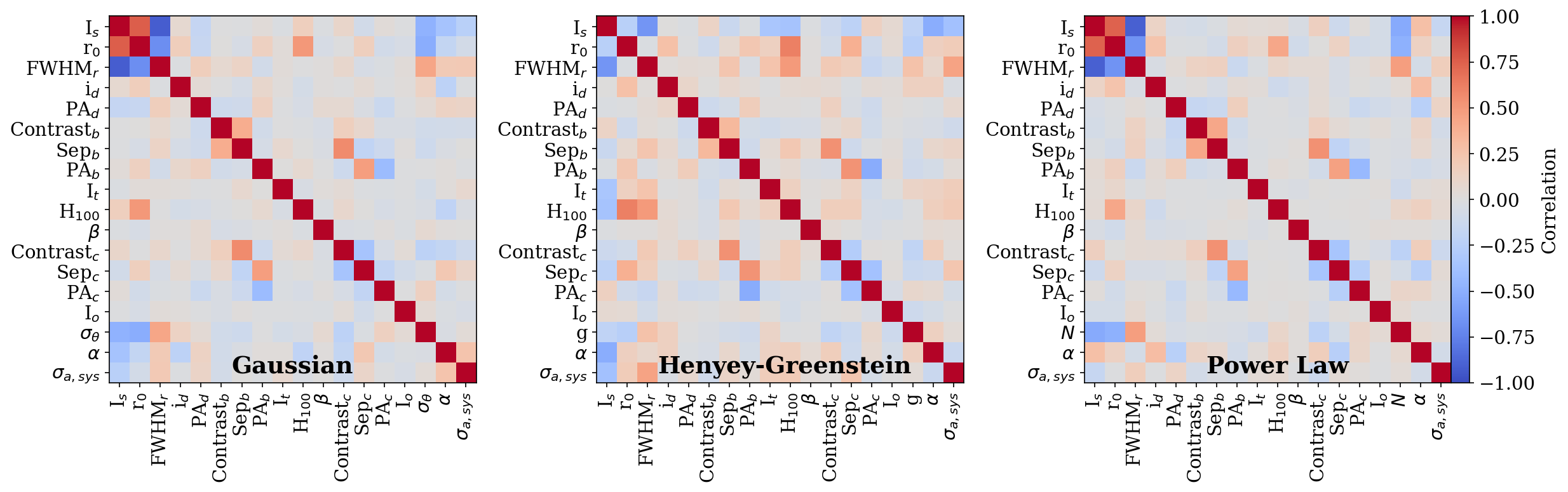}
\caption{The correlation matrices, from left to right, of the model parameters for the Gaussian model, the Henyey-Greenstein model and the power law model.}
\label{fig:f480M_correlations}
\end{figure*}

{Figure \ref{fig:f480M_corner} shows a corner plot of the planet parameters for each of the three models, as well as the Bayesian average of the three models. We note that there are correlations between the parameters of both planets. Notably, there is a moderate anti-correlation between the separation and the contrast of PDS 70 c. From the correlation matrices of the model parameters, shown in Figure \ref{fig:f480M_correlations}, there are no strong correlations between the disk parameters and the planet contrasts, indicating that the disk model has a minimal effect on the measured contrasts.}

\begin{deluxetable}{ccc}
    \label{tab:comp_fit_par_wsys}
    \tablecaption{PDS 70 b and c parameters}
    
    \tablewidth{0pt}
    
    \tablehead{& PDS 70 b & PDS 70 c }
    \startdata
    \hline
     {Separation (mas)}&$150.5^{+9.5}_{-10.9}$ & $218.4^{+7.2}_{-8.2}$ 
     \\
     {Position Angle ($^\circ$)}&$131.2^{+1.5}_{-1.6}$ & $270.0^{+1.7}_{-1.7}$
     \\
     {Contrast ($\Delta$mag)}&$5.84^{+0.07}_{-0.07}$&$6.48^{+0.17}_{-0.14}$ 
     \\
     {Flux$^1$ ($\mu$Jy)}& $629.7^{+42.2}_{-39.0}$ &$349.5^{+48.5}_{-49.6}$
     \\
     \hline
     \enddata
     \footnotesize{Notes: We report the median along with the 16th and 84th percentiles of the Bayesian average of the marginalized posteriors over the three disk (Equations \ref{eqn:odazimuthal}, \ref{eqn:odazimuthal_Gaussian}, \ref{eqn:odazimuthal_powerlaw}) plus plus star planets models we explored.
     
     $^1$ We convert the planet contrasts to flux using our measured star plus unresolved inner disk flux of $136.4 \pm 6.8$ mJy.
     }
\end{deluxetable}

\subsection{Residual Signal and Contrast Limits}

\begin{figure}
\centering
\includegraphics[width=1\linewidth]{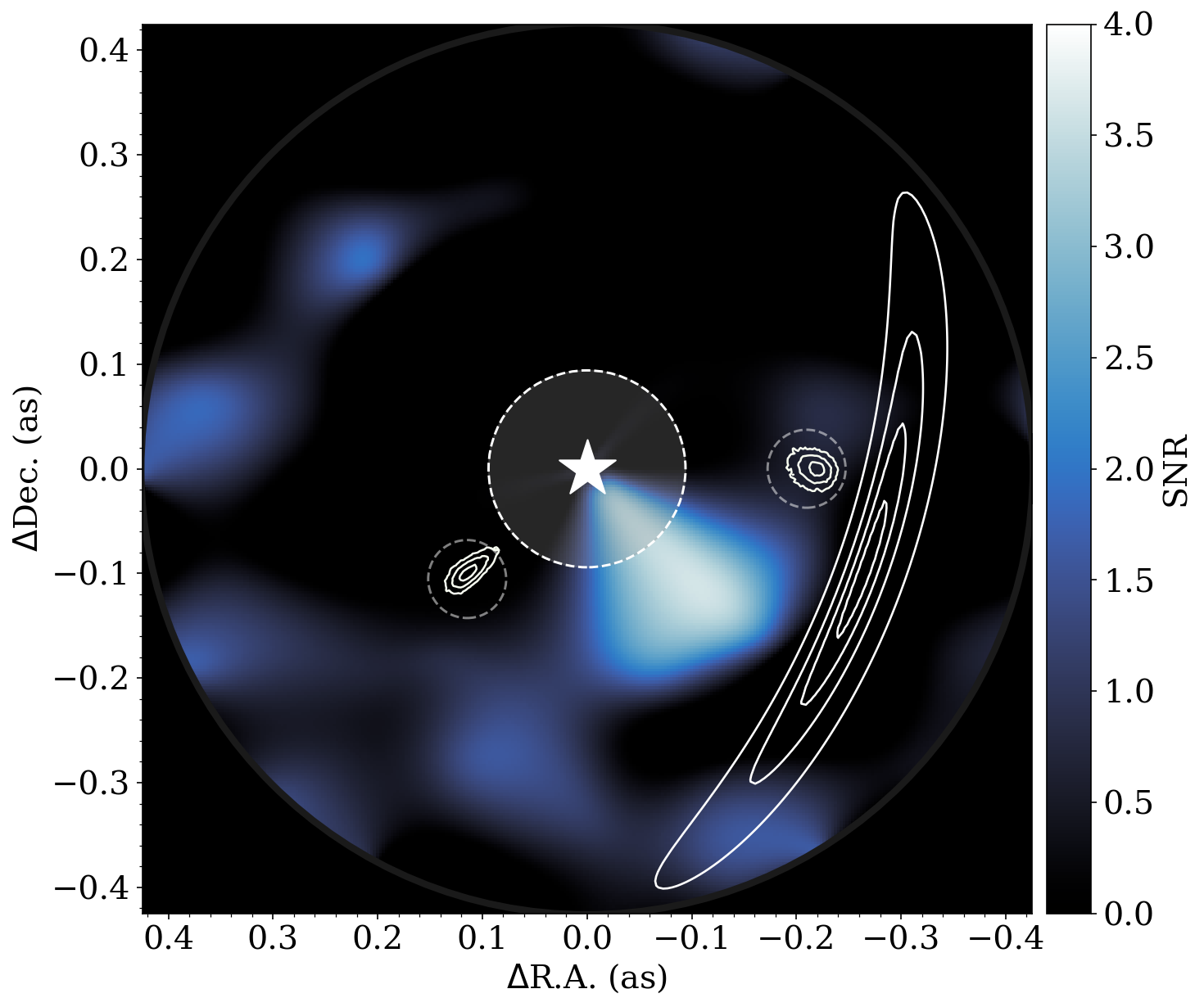}
\caption{Point source model SNR map, {within a radius of 400 mas (denoted by the grey circle)}, calculated from the {kernel} phase residuals from the two point sources plus {power law} geometrical model fit. The white contours denote the 1, 2 and 3 $\sigma$ contours from the posterior calculated with nested sampling. The grey dashed lines denote the predicted locations of PDS 70 b and c at the time of the observations \citep{2021AJ....161..148W,2021ascl.soft01003W}. The white star denotes the position of the star. The extended white contours to the south west of the star denote the forward scattering side of the outer disk  {from the power law} model. As in Figure \ref{fig:f480M_fit}, the white dashed circle denotes the diffraction limit of the data.}

\label{fig:f480M_2pt_residuals_SNR}
\end{figure}

We compute an SNR map, shown in Figure \ref{fig:f480M_2pt_residuals_SNR}, made by dividing the best fit point source binary contrast to the residual {kernel phase} 
signal by its associated uncertainty fitting.

{Because closure phases are non-linear, we calculate the complex visibilities of the {power law model}, using the median of the marginalized posterior for all parameters of {the power law model, } 
and add this to our binary model visibilities before calculating the model closure phases {and then kernel phases}. Each cell in the grid shows the best fit contrast divided by the uncertainty in the contrast calculated using the Laplace approximation \citep{doi:10.1080/01621459.1986.10478240}, assuming the uncertainties are Gaussian. We clearly see that there is residual emission at an SNR of $\sim$4, to the south west of PDS 70 A and very close to the inner working angle of 94 mas, that is not fully captured by our two planet plus disk model. To assess the nature of the signal that we observe, we use nested sampling to fit a star plus planet model to the residual {kernel phases and find a contrast of $7.6^{+0.8}_{-1.3}$ magnitudes, a separation of $118^{+52}_{-64}$ mas and a position angle of $220^{+10}_{-15}$ degrees}. The position angle of the signal is {reasonably well constrained, however, there is a strong correlation} between the separation and contrast, which is expected for a high contrast source near the diffraction limit. Interestingly, the position angle of the observed signal is not consistent with forward scattering from an inner disk that has the same geometry as the outer disk. However, due to the unconstrained nature of the separation of the emission it is unclear whether the signal is due to compact emission or an inner disk feature.}

{To assess the limits that we are able to place on the presence of additional planets in the system, we calculate a 5$\sigma$ contrast curve, from the residual {kernel} phase signal of the power law model. {We use the approach described above to fit to the residual {kernel phase} 
signal, and calculate 
Bayesian upper limits. To do this, we adapt the method presented by \cite{2018AJ....156..196R} to interferometric data}. We first calculate a grid of likelihoods for a binary model at fixed R.A.\ and Dec.\ locations for a range of contrasts. We next use the maximum likelihood point in each grid cell to initialize gradient descent, using the BFGS algorithm \citep{10.1093/imamat/6.1.76,10.1093/comjnl/13.3.317,35d0019d-775a-3628-b0b4-67be112e346b,e3177091-3094-3792-9d61-0ab445735ddb}, to find the contrast that maximizes the likelihood. We use the calculated contrast in each cell, along with the Laplace approximation to calculate the uncertainty in the calculated contrast at each R.A. and Dec.\ point. Finally, {combining the calculated contrasts, uncertainties and a 99.999971\% (equivalent to a $5\sigma$) cutoff probability and using} Equation 8 in \cite{2018AJ....156..196R}, {we} calculate the {5$\sigma$} upper limit at each point {within a radius of 400 mas, which we} azimuthally average over to generate a contrast curve. We henceforth refer to this method as the Ruffio method. }

\begin{figure*}
\centering
\includegraphics[width=0.9\linewidth]{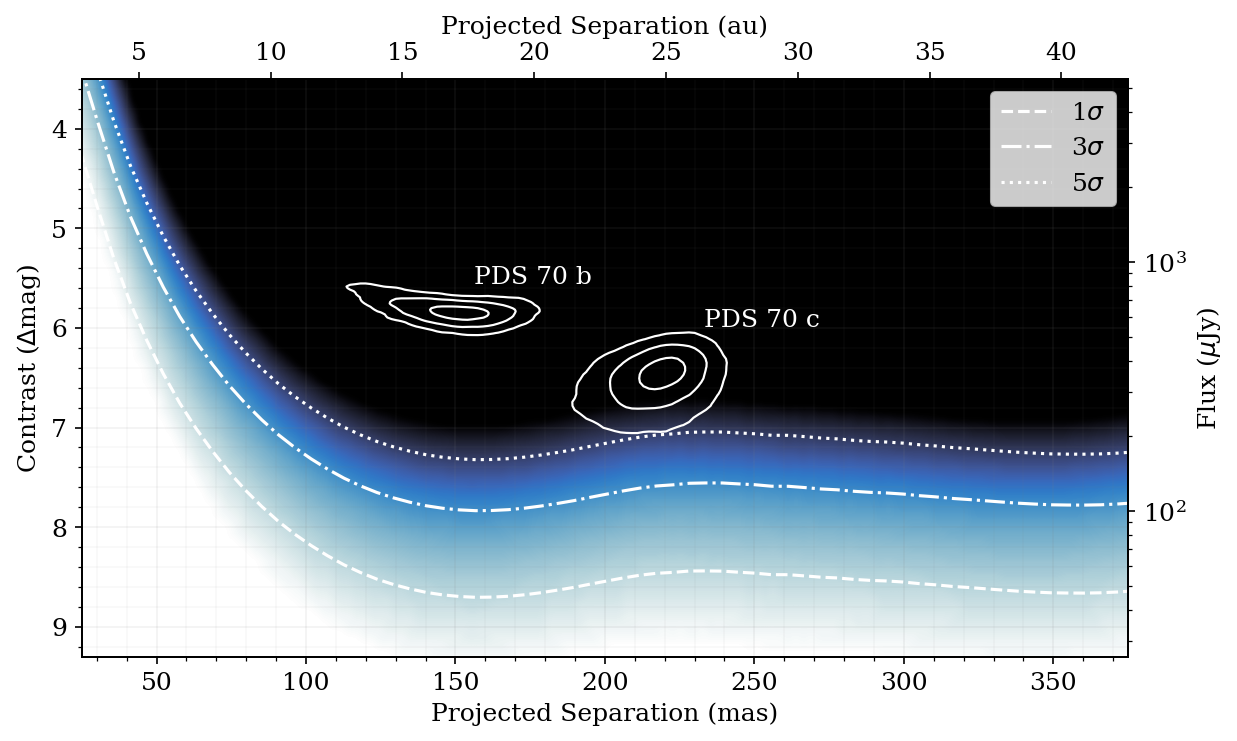}
\caption{Mean contrast curves calculated from the residuals of the {power law} geometrical model plus two point source fit. {The colour map represents the 0-7$\sigma$ contrast upper limits, linearly from white to black. We also denote the 1, 3 and 5$\sigma$ azimuthally averaged limits with the white curves.} 
The 1, 2, and 3 $\sigma$ confidence intervals of the planet contrast/separation, calculated from the nested sampling posterior of the {power law model fit, are denoted by the white contours}. {We note that, for the signal depth of our data, the expected {achievable contrast} for photon noise dominated data is $\sim$7.8, by Equation 10 in \cite{2023PASP..135a5003S}.}}
\label{fig:f480M_cont_cp_corr_im_sub}
\end{figure*}

The derived 5$\sigma$ contrast/upper limit curves{, along with the 1 and 3$\sigma$ curves, 
calculated using the {
Ruffio method}, are shown in Figure \ref{fig:f480M_cont_cp_corr_im_sub}. Beyond 110 mas, we reach a 5$\sigma$ contrast upper limit of $>$7 magnitudes, and a 3$\sigma$ limit of $\sim$7.6-7.8 magnitudes, which is consistent with the expected photon noise limit of $\sim$7.8, given by Equation 10 by \cite{2023PASP..135a5003S}.} The contours show the 1, 2 and 3 $\sigma$ confidence intervals on the contrast and the separation for PDS 70 b and c calculated from the {Bayesian average of the three model posteriors}.

We also calculate a 5$\sigma$ upper limit on any flux at the reported position of the planet candidate PDS 70 d from \cite{2024arXiv240304855C} (sep.\ = 115.2 mas, P.A. = 291.8$^{\circ}$), as our observations are only separated by 12 days. We first calculate the contrast posterior distribution using \texttt{dynesty}, with uniform priors on the contrast, from -1 to 1. {We find a mean contrast of $-0.1$ $(\pm$ $3.0)$ $\times$ $10^{-4}$. To convert this measured contrast to an upper limit, we use Equation 8 by \cite{2018AJ....156..196R}, with the measured mean, standard deviation and a cutoff probability of 99.999971\%, which is equivalent to a 5$\sigma$ upper limit, with a positivity prior on the contrast. From this, we calculate a 5$\sigma$ upper limit of 7.04 mag, corresponding to a flux of 208 $\pm$ 10 $\mu$Jy. We also calculate the 5$\sigma$ upper limit using the Laplace approximation, again retrieving an upper limit of 7.04 mag, in line with our assumption of a Gaussian posterior}.

\subsection{SED Fitting}
\label{sec:sed}

We fit the available data points of both protoplanets including the new NIRISS/AMI
F480M measurement. For planet b, we use the SPHERE/IFS spectrum and SPHERE/IRDIS photometry from \cite{2018A&A...617L...2M}, the VLTI/GRAVITY spectrum from \cite{2021AJ....161..148W}, the VLT/NaCo {photometry} at $3-5~\mu$m from \cite{2020A&A...644A..13S}, and the F187N and F480M NIRCam measurements from \cite{2024arXiv240304855C}. For planet c, we consider the SPHERE/IRDIS and VLT/NaCo {photometry} from \cite{2020A&A...644A..13S}, the SPHERE/IFS and VLTI/GRAVITY from \cite{2021AJ....161..148W}, and the F187N and F480M NIRCam measurements from \cite{2024arXiv240304855C}.

Given the best-fit presented by \cite{2021AJ....161..148W}, we use the Drift-PHOENIX models \citep{2003A&A...399..297W, 2004A&A...414..335W, 2006A&A...455..325H, 2008ApJ...675L.105H} to describe the atmospheric emission from the protoplanets. Following \cite{2021AJ....161..148W}, we do not apply any correction for extinction. The parameter space is explored using {\tt pymultinest} \citep{2014A&A...564A.125B}, employing 1000 live points. For both planets, we first fit the available data without including any CPD contribution. We then include a CPD contribution in the form of a single temperature blackbody. Blackbody emission has been shown to be the simplest model able to describe well the circumplanetary disk emission of GQ~Lup~B up to $11.7~\mu$m \citep{2024arXiv240407086C}. Although a blackbody is a simple model and it is not the only possible solution to characterize CPD emission, more complex models would not provide useful information as the CPD contribution is mostly determined by the F480M photometry. {The model parameters for both PDS 70 b and c are shown in Section \ref{sec:appsedfit}.}

\begin{figure*}
\centering
\includegraphics[width=0.9\linewidth]{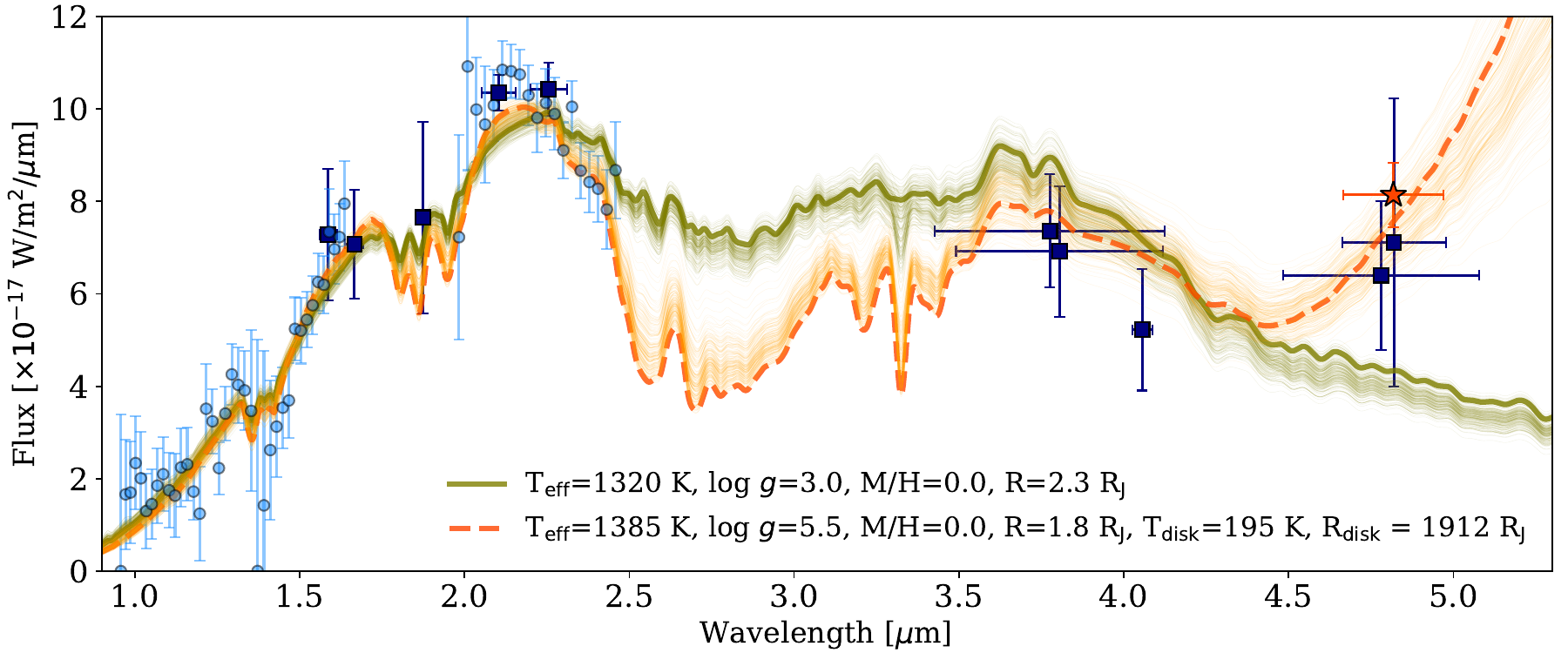}
\caption{Spectral fit of PDS70 b using Drift-PHOENIX models with (dashed orange line) and without (solid green line) contribution from a CPD. Blue circles represent the IFS and GRAVITY spectra, 
while the photometric datapoints are reported with a blue square {(see Section \ref{sec:sed} for details)}. Horizontal errorbars represent the effective width of the filters. The red star shows the new NIRISS F480M measurement. The thick lines are the spectra obtained from the set of parameters providing the maximum likelihood, while the thin lines report 100 samples randomly drawn from the posterior distribution.}
\label{fig:spectrum_b}
\end{figure*}

For planet b, the best fit without a CPD component is reported by the green solid line in Figure~\ref{fig:spectrum_b}. The spectrum reproduces well the IFS spectrum, the SPHERE $H$ band {photometry}, and the F187N measurement obtained with NIRCam. However, the first half of the GRAVITY spectrum is underestimated, while the second appears to be overestimated. Finally, at $\lambda>3~\mu$m, almost none of the data points appear to agree within $1\sigma$ with the best-fit model, with the exception of the NIRCam F480M photometry, which has a very large uncertainty. When including the CPD contribution, the second half of the GRAVIITY spectrum is better reproduced, while at longer wavelengths the predicted flux is closer to the measurements (see orange dashed line in Figure~\ref{fig:spectrum_b}). In terms of planet model parameters, the presence of MIR excess emission in the modeling suggests a hotter and smaller planet. 
{The Bayes factor comparing the blackbody model to the no blackbody model is $2.8 \times 10^{10}$. Hence, the presence of the additional blackbody emission provides a better description of the observations. }

\begin{figure*}
\centering
\includegraphics[width=0.9\linewidth]{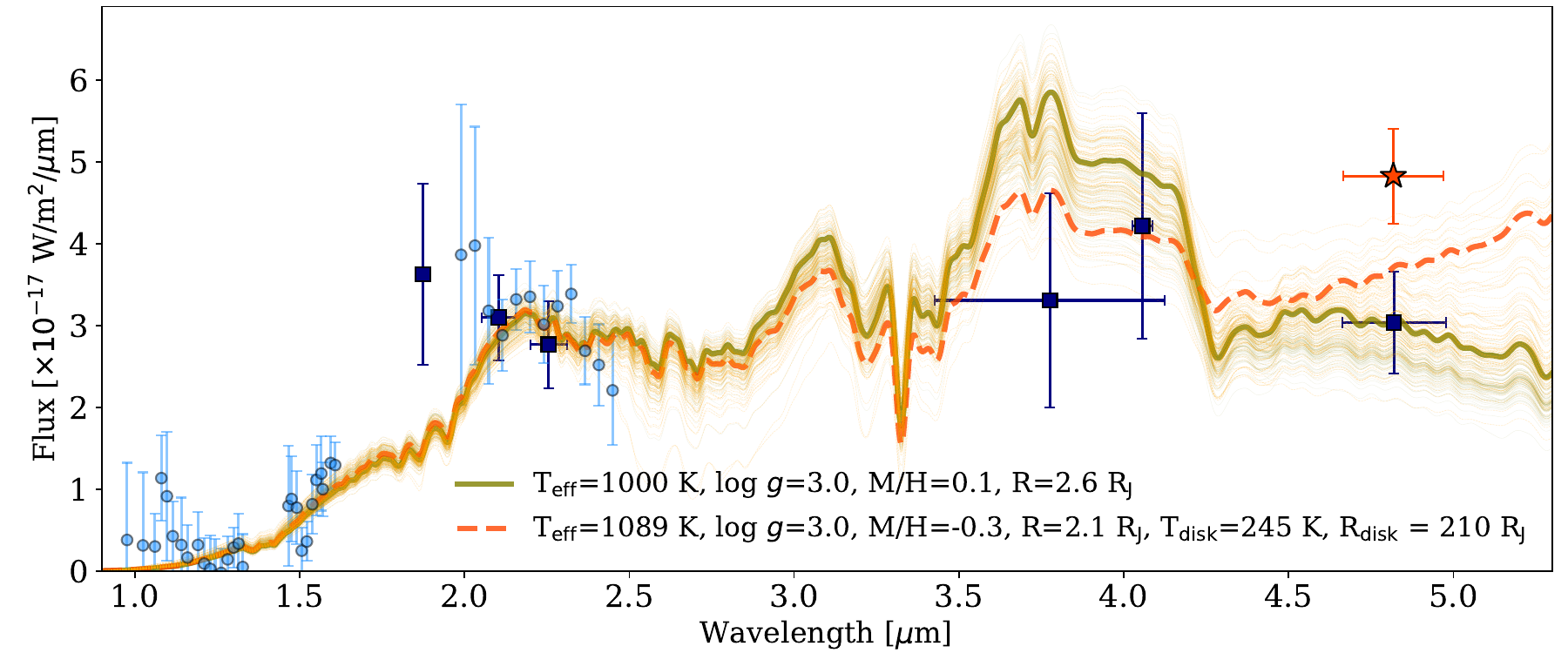}
\caption{Same as Figure~\ref{fig:spectrum_b}, for PDS 70 c.}
\label{fig:spectrum_c}
\end{figure*}

The results for planet c are displayed in Figure~\ref{fig:spectrum_c}. As reported by \cite{2024arXiv240304855C}, the F187N shows an excess emission, possibly due to Pa$\alpha$ emission. The blue region of the GRAVITY spectrum does not agree with the fit, and the $L'$ measurement is off by almost $2\sigma$. Also, our NIRISS/AMI F480M measurement is clearly at odds with respect to the green solid line model fit, which does not include a CPD component. The spectrum of the best model once including the CPD presents a lower $L'$ flux, consistent with the NaCo measurement. Furthermore, the CPD contribution passes right in between the F480M measurements from NIRCam and NIRISS. {The Bayes factor of the model including the blackbody compared with the blackbody-free model is 0.88, indicating that the model without the additional blackbody component is preferred. However, we note that our measurement of PDS 70 c hints that there is in fact excess emission that can't be explained by the atmosphere only model, and the fit is being constrained by the fainter measurement of PDS 70 c made by \cite{2024arXiv240304855C}. Follow-up observations at 4.8 $\mu$m with NIRISS and NIRCam, and beyond 5 $\mu$m with MIRI, could be used to confirm the presence of warm CPD emission and constrain the CPD properties.}  






    
    
     



\section{Discussion} 
\label{sec:disc}

\subsection{PDS b and c F480M Photometry}

{For PDS 70 b, we find a F480M flux of $629.7^{+42.2}_{-39.0}$ $\mu$Jy. Similarly, for PDS 70 c, we find a flux of $349.5^{+48.5}_{-49.6}$ $\mu$Jy. While our result for PDS 70 b is consistent with the NIRCam F480M results of $553.5 \pm 241.4$ $\mu$Jy published by \cite{2024arXiv240304855C}, 
the PDS 70 c value is significantly inconsistent with the NIRCam F480M results of $236.5 \pm 48.3$ $\mu$Jy.}


There are several possible reasons for the PDS 70 c discrepancy between NIRCam and NIRISS/AMI. It is possible that the {fluxes published by \cite{2024arXiv240304855C}} are systematically underestimated 
due to self/over subtraction, as a result of the small roll angle of $\sim$5 degrees between their two observations. Another possibility for an underestimated brightness of PDS 70 c could be due to over subtraction of the spatially coincident disk. {\cite{2024arXiv240304855C}} use a radiative transfer model based on models published by \cite{2022A&A...658A..89P} and \cite{2023A&A...677A..76P}, that were originally fit to ALMA data and 1.25 $\mu$m polarized VLT/SPHERE data of PDS 70. These models were iterated on by fitting for the minimum grain size and settling parameters to less than half of the forward scattering side of the disk, in F480M. This tightly constrained fitting approach could lead to biased results that do not adequately reproduce the disk emission at the location of PDS 70 c. Specifically, an incorrect estimation of the scattering phase function will significantly impact the PDS 70 c measured contrast. {It is also possible that the flux of PDS 70 presented in this work is over estimated, due to emission from the spiral structure that was tentatively detected in \cite{2024arXiv240304855C}, that is not included in our geometrical model. However, any bias from not including this component in our model should be minimal, due to its extended nature. A final explanation for the significant difference between our flux measurements of PDS 70 c is that the flux of PDS 70 c itself is variable at 4.8 $\mu$m, due to non-steady accretion from the circumplanetary material.}

{Separating out the disk emission is less of an issue for the 
{interferometric} data set despite the simple disk {models employed, using a symmetric, optically thick disk component plus the {asymmetric} Gaussian, power law and Henyey-Greenstein azimuthal brightness profiles.} 
The signals of the disk and the planet in Fourier space are well separated and thus we can disentangle the asymmetric contributions from the forward scattering peak of the disk, and PDS 70 c, offset by $\sim$20 degrees.
Furthermore, our joint modelling of the planet and disk parameters allows us to measure the correlations between the parameters of both planets and the disk, which are shown by the {correlation matrices in Figure \ref{fig:f480M_correlations}}. {Notably for the contrast of PDS 70 c, there are moderate correlations with the model parameters that control the azimuthal brightness profile of the disk, $\sigma_{\theta}$, g and $N$ (more forward scattering leads to less emission), a strong correlation with the separation of PDS 70 b (larger separation leads to less emission), a moderate anti-correlation with $\alpha$, the disk inner-edge truncation parameter, and a moderate anti-correlation with its own separation from the star (plus unresolved inner disk emission). }}

{We note that these correlations, between the contrast of PDS 70 c and the disk model parameters, are small compared to the correlation between the contrast of PDS 70 c and the separation of PDS 70 b, which is evident in {Figure \ref{fig:f480M_correlations}}. This means a larger separation of PDS 70 b would correspond to a fainter contrast of PDS 70 c. To asses the impact of this correlation, we calculate the contrast of PDS 70 c using only posterior samples within 3$\sigma$ of the predicted location of PDS 70 c from the orbit fitting presented in \cite{2021AJ....161..148W}. We find a contrast of $6.54^{+0.15}_{-0.14}$ mags, consistent with our value of $6.48^{+0.17}_{-0.14}$ mags, calculated using the full posterior.} 


{As another test, we briefly explored including an inner disk component in our {models}, with its geometry fixed to that of the outer disk. We tested symmetric models as well as models that included skewed emission in the forward scattering direction. We found that symmetric models did not noticeably improve the fit. For the skewed models, we found that the fit did somewhat improve, for a large, bright, extended inner disk model, however, we saw significant correlations between the inner disk and the planet location parameters. {Such a bright and large inner disk is inconsistent with the size and mass of the inner disk that has been inferred from SED fitting \citep[e.g.,][]{2024arXiv240309970G}.} {Additionally, the planet contrasts that we found from these tests were all consistent with the planets plus outer disk only model, well within $1\sigma$ of the contrasts that we report in Table \ref{tab:comp_fit_par}.} Given these points, as well as the fact that the inner disk is marginally resolved, if at all, and that it possibly has a complicated morphology \citep{2019A&A...632A..25M,2022MNRAS.513.5790C}, not well captured by a geometrical model, we decided not to further explore an inner disk model component.} 




\subsection{Contrast Limits and the Nature of the Signal Seen in the Residuals}

The residual signal shown in Figure \ref{fig:f480M_2pt_residuals_SNR}, is in a direction significantly offset from the expected forward scattering direction of the inner disk. This indicates that what we observe is not due to a simple inner disk structure, and may hint at a complex inner disk morphology such as a spiral or clumpy features, as has been suggested by \cite{2022MNRAS.513.5790C}, from ALMA observations in the mm, and by \cite{2019A&A...632A..25M} from near-infrared observations with VLT/SPHERE. {It is also possible that the asymmetry we observe is related to the larger separation, $\sim$180 mas, ALMA compact emission seen by \cite{2023A&A...675A.172B}, as it is at very similar position angle.} {It might even highlight extended emission from an accretion stream between PDS 70 b and c, tentatively detected by \cite{2024arXiv240304855C}.} 
Another scenario is that the signal we observe is due to an additional planet interior to the orbit of PDS 70 b. 
To distinguish between these {many} scenarios, follow-up observations, at similar wavelengths, will be necessary with {NIRISS/AMI.} 
Analysing follow-up data using an approach similar to \cite{2023AJ....165...29T}, using \texttt{Octofitter}\footnote{\url{https://sefffal.github.io/Octofitter.jl/dev/}} \citep{2023AJ....166..164T}, would allow for a robust detection of any orbital motion, even with low {significance} detections or non-detections at individual epochs. 

Additionally, it is clear from our SNR maps (Figure \ref{fig:f480M_2pt_residuals_SNR}), and our contrast upper limit of {7.04} mags derived in Section \ref{sec:res}, that we do not detect a signal (to the north-west of PDS 70 A, at a P.A. $\approx$ 291.8$^{\circ}$) consistent with the point-like feature (PLF) seen by \cite{2019A&A...632A..25M}/PDS 70 d seen by \cite{2024arXiv240304855C}, detected at shorter wavelengths. 
This indicates that if the signal is a planet, it likely traces bright scattered light emission from a potential planetary envelope instead of {indicating another}  warm source like both PDS 70 b and c. This scenario would be similar to the protoplanet candidate HD 169142 b \citep{2023MNRAS.522L..51H}.

Our contrast upper limits, shown in Figure \ref{fig:f480M_cont_cp_corr_im_sub}, are the deepest limits on additional point source emission within $\sim$250 mas in the disk gap of PDS 70, at 4.8 $\mu$m. This clearly demonstrates the power of {NIRISS/AMI} 
at probing small angular scales compared with direct imaging that achieves significantly better contrasts at larger separations, as is seen by \cite{2024arXiv240304855C} {\citep[see also][]{2023ApJ...951L..20C}}. 


\section{Conclusions} 
\label{sec:con}

In this work, we present James Webb Interferometer observations of PDS 70 with the NIRISS F480M filter, the first space-based interferometric observations of this system. Using a joint model fitting approach to simultaneously fit for star, disk, and planet emission, we re-detect the protoplanets PDS 70 b and c, and derive fluxes of both planets in this filter (Table \ref{tab:comp_fit_par_wsys}). Additionally, we place the deepest constraints on additional planets within the disk gap of PDS 70 inside $\sim$250 mas in F480M, and calculate an F480M upper limit on the flux of the candidate PDS 70 d. We also detect a new feature at an SNR of $\sim$4, to the south of PDS 70 A, whose nature is uncertain and will require follow-up observations to confirm. 

Furthermore, our results show that NIRISS/AMI can reliably measure relative astrometry and contrasts of young planets in a part of parameter space (small separations and moderate to high contrasts) that is unique to this observing mode, and inaccessible to all other present facilities at 4.8 $\mu$m.
We demonstrate a NIRISS/AMI observing strategy for targets faint enough to acquire greater than $\sim$10 groups up the ramp before the signal in the central 9 pixels reaches a total (linearized) value of 30,000 DN ($\sim$48,000 electrons). {We show that by using this stringent data selection criteria, we achieve nearly photon noise limited performance.}
{For observing brighter targets, and to overcome} the limitation of the sparsity of the NIRISS AMI uv-coverage, methods that are able to analyze the data directly in the image plane (or using the full extent of the Fourier ``splodges") will be necessary. A promising approach will be forward modelling of the full optical system and detector systematics (e.g., the brighter-fatter effect, 1/f noise) as is made possible with $\partial$Lux \citep{Desdoigts2023DifferentiableOW}.


\section{acknowledgements}

{The authors would like to thank the referee for their comments and suggestions that significantly strengthened the paper. The authors also thank Gregory Herczeg for his helpful comments.} 

This work is based on observations made with the NASA/ESA/CSA James Webb Space Telescope. The data were obtained from the Mikulski Archive for Space Telescopes at the Space Telescope Science Institute, which is operated by the Association of Universities for Research in Astronomy, Inc., under NASA contract NAS 5-03127 for JWST. These observations are associated with program GTO 1242, and can be accessed via doi:\href{https://doi.org/10.17909/6qvy-zr60}{10.17909/6qvy-zr60}.

D.B. and D.J. acknowledge the support of the Natural Sciences and Engineering Research Council of Canada (NSERC). G.C. thanks the Swiss National Science Foundation for financial support under grant numbers P500PT\_206785 and P5R5PT\_225479. M.D.F. is supported by an NSF Astronomy and Astrophysics Postdoctoral Fellowship under award AST-2303911. J.S.-B. acknowledges the support received from the UNAM PAPIIT project IA 105023.

We acknowledge and respect the L\textschwa\'{k}\textsuperscript{w}\textschwa\textipa{\ng}\textschwa n (Songhees and Esquimalt) Peoples on whose territory the University of Victoria stands, and the L\textschwa\'{k}\textsuperscript{w}\textschwa\textipa{\ng}\textschwa n and \underline{W}SÁNEĆ Peoples whose historical relationships with the land continue to this day.

BJSP acknowledges the traditional owners of the land on which the University of Queensland is situated, and PGT and LD would like to acknowledge the Gadigal People of the Eora nation: upon whose unceded, sovereign, ancestral lands they works, and pays respect to their Ancestors and descendants, who continue cultural and spiritual connections to Country.

\noindent 

\software{\texttt{JAX} \citep{jax2018github}, \texttt{NumPy} \citep{harris2020array}, \texttt{dynesty} \citep{2020MNRAS.493.3132S,sergey_koposov_2023_7600689}, \texttt{Astropy} \citep{2013A&A...558A..33A,2018AJ....156..123A,2022ApJ...935..167A}, \texttt{Matplotlib} \citep{Hunter:2007}, \texttt{corner} \citep{corner}, {\tt pymultinest} \citep{2014A&A...564A.125B}, \texttt{jwst} \citep{bushouse_2023_8157276}.}

\bibliography{sample631}{}
\bibliographystyle{aasjournal}

\appendix



\section{Model Fitting Results}
\label{sec:desc_of_geo}

\subsection{Disk Parameters}

{Table \ref{tab:disk_par} shows the derived disk parameters for each of the Henyey-Greenstein, Gaussian and power law models. 
In all 
cases}, we used Gaussian priors on the disk geometry, with the mean values taken from \cite{kep1} and \cite{kep2} and sensibly small standard deviation values chosen so as to break degeneracies between the disk model parameters, using the previously measured geometry of the disk. This was done because we are not primarily interested in independently measuring the parameters of the disk but wish to produce a sensible model of the disk emission so as to measure the emission from the known planets. 

The derived disk parameters are nearly identical between the models. 
This confirms what we see in the correlation plot in Figure \ref{fig:f480M_corner}, that there are no significant correlations between the disk parameters and the planet parameters {that could be biasing our results}. 
If {strong correlations were present} 
we would see significant differences between the model disk parameters between the two cases.

\begin{deluxetable*}{ccccc}
    \label{tab:disk_par}
    \tablecaption{PDS 70 disk model parameters}
    
    \tablewidth{0pt}
    
    \tablehead{Parameters&Prior&Henyey-Greenstein& Gaussian &Power Law}
    \startdata
    \hline
     {$\log{A_a}$ (arb.)}& $\mathcal{U}(-10,-4)$& $-4.53^{+0.09}_{-0.09}$ & $-5.11^{+0.12}_{-0.10}$ &  $-5.14^{+0.11}_{-0.10}$\\
     {$r_0$ (as)}& $\mathcal{U}(0.35,0.65)$& $0.44^{+0.02}_{-0.02}$ & $0.46^{+0.04}_{-0.03}$ & $0.46^{+0.03}_{-0.03}$\\
     {FWHM$_r$ (as)}& $\mathcal{U}(0.03,1.00)$& $0.29^{+0.08}_{-0.06}$ & $0.22^{+0.06}_{-0.05}$ & $0.22^{+0.06}_{-0.05}$\\
     {$i$ ($^{\circ}$)}& $\mathcal{N}(51.7,1.0)^1$& $52.7^{+0.9}_{-1.0}$ & $52.2^{+0.9}_{-0.9}$ & $52.3^{+0.9}_{-0.9}$\\
     {P.A. ($^{\circ}$)}& $\mathcal{N}(160.4,1.0)^1$& $160.0^{+0.8}_{-0.9}$ & $160.2^{+0.9}_{-0.8}$ & $160.4^{+0.9}_{-0.9}$\\
     {$\log{A_s}$ (arb.)}& $\mathcal{U}(-10,-4)$& $-8.6^{+1.1}_{-1.0}$ & $-8.2^{+1.2}_{-1.2}$ & $-8.2^{+1.1}_{-1.2}$ \\
     {$H_{100}$ (au)}& $\mathcal{N}(13,5)^{2,3}$& $9.9^{+5.3}_{-4.7}$ & $13.7^{+4.9}_{-4.7}$ & $14.7^{+4.5}_{-4.8}$ \\
     {$\beta$ }& $\mathcal{N}(1.25,0.01)^2$& $1.25^{+0.01}_{-0.01}$ & $1.25^{+0.01}_{-0.01}$ &$1.25^{+0.01}_{-0.01}$ \\
     {$\log{I_o}$ (arb.)}& $\mathcal{U}(-10,-2)$& $-6.7^{+3.0}_{-2.2}$ & $-6.2^{+2.8}_{-2.6}$ &$-5.9^{+2.6}_{-2.7}$ \\
     {$\alpha$}& $\mathcal{U}(0,10)$& $5.4^{+3.2}_{-3.5}$ & $2.5^{+4.9}_{-1.9}$ & $2.7^{+4.4}_{-2.0}$\\
     {$\sigma_{a,sys}$}& log$\mathcal{U}(0.0001,0.1)$& $0.009^{+0.003}_{-0.002}$ & $0.006^{+0.002}_{-0.001}$ & $0.006^{+0.002}_{-0.001}$\\
     {$g$}& $\mathcal{U}(0,1)$& $0.33^{+0.04}_{-0.04}$ & -- & -- \\
     {$\sigma_{\theta}$ ($^{\circ}$)}& $\mathcal{U}(60,360)$& -- & $106^{+12}_{-9}$ & \\
     {$N$}& $\mathcal{U}(0,100)$& -- & -- & $6.6^{+1.2}_{-1.2}$ \\
     {$\log{\mathcal{Z}}^4$} & -- & $-31.3 \pm 0.1$& $-25.0 \pm 0.1$ & $-24.7 \pm 0.1$ \\
     \hline
     \enddata
     \footnotesize{Notes: The median along with the 16th and 84th percentiles of the marginalized posteriors calculated with nested sampling are reported. $\mathcal{U}(a,b)$ denotes a uniform prior with lower and upper bounds a and b, respectively. $\mathcal{N}(\mu,\sigma)$ denotes a Gaussian prior, with mean $\mu$ and standard deviation $\sigma$. 
     
     $^1$ Reference: \cite{kep2}.
     
     $^2$ Reference: \cite{kep1}, using the same distance as in the original paper of 113.43 pc \citep{gai18}.
     
     $^3$ Truncated to be positive.
     
     $^4$ The logarithm of the Bayesian evidence for each model.}
     
\end{deluxetable*}

\subsection{Planet Parameters}

In Table \ref{tab:comp_fit_par} we show the planet parameters, calculated using nested sampling, from each of the Henyey-Greenstein, Gaussian and power law models. 

\subsection{SED Fit Parameters}
\label{sec:appsedfit}
Table \ref{tab:sed_fit} shows the Drift-PHOENIX model parameters, with and without a blackbody component to model any excess CPD emission. We also report the Bayes factor, B$_{12}$, comparing each model with a blackbody component to the model without a blackbody component.

\begin{deluxetable*}{ccc}
    \label{tab:comp_fit_par}
    \tablecaption{PDS 70 b and c parameters from individual models}
    
    \tablewidth{0pt}
    
    \tablehead{& PDS 70 b & PDS 70 b }
    \startdata
    \hline
     Henyey-Greenstein \\
     \hline
     {Separation (mas)}&$152.4^{+9.0}_{-10.2}$ & $215.7^{+7.8}_{-8.9}$ 
     \\
     {Position Angle ($^\circ$)}&$130.2^{+1.6}_{-1.8}$ & $270.6^{+1.8}_{-1.7}$
     \\
     {Contrast ($\Delta$mag)}&$5.85^{+0.07}_{-0.07}$&$6.46^{+0.16}_{-0.14}$ 
     \\
     \hline
     Gaussian \\
     \hline
     {Separation (mas)}&$150.7^{+9.4}_{-10.8}$ & $218.7^{+7.1}_{-8.0}$ 
     \\
     {Position Angle ($^\circ$)}&$131.1^{+1.5}_{-1.6}$ & $270.0^{+1.7}_{-1.7}$
     \\
     {Contrast ($\Delta$mag)}&$5.84^{+0.07}_{-0.07}$&$6.47^{+0.17}_{-0.14}$ 
     \\
     \hline
     Power Law \\
     \hline
     {Separation (mas)}&$150.3^{+9.6}_{-10.8}$ & $218.3^{+7.2}_{-8.3}$ 
     \\
     {Position Angle ($^\circ$)}&$131.3^{+1.5}_{-1.5}$ & $270.0^{+1.8}_{-1.7}$
     \\
     {Contrast ($\Delta$mag)}&$5.84^{+0.07}_{-0.07}$&$6.48^{+0.17}_{-0.14}$ 
     \\
     \enddata
     \footnotesize{Notes: We report the median along with the 16th and 84th percentiles of the marginalized posteriors of each model.
     
     }
\end{deluxetable*}

\begin{deluxetable*}{cccccccc}
    \label{tab:sed_fit}
    \tablecaption{Drift-PHOENIX SED model parameters for PDS 70 b and c}
    
    \tablewidth{0pt}
    
    \tablehead{&T$_{\mathrm{eff}}$& $\log g$ &[M/H] & R  &T$_{\mathrm{disk}}$& R$_{\mathrm{disk}}$ & B$_{12}$ \\
     \nocolhead{name}&\colhead{(K)}&\nocolhead{name}&\nocolhead{name}&\colhead{(R$_{J}$)}&\colhead{(K)}&\colhead{(R$_{J}$)}& \nocolhead{name}}
    \startdata
    \hline
     {PDS 70 b}& $1331^{+15}_{-10}$& $3.14^{+0.81}_{-0.11}$& $0.02^{+0.13}_{-0.16}$ & $2.22^{+0.06}_{-0.07}$&$-$& $-$& $-$ \\
   {}& $1402^{+27}_{-22}$& $5.37^{+0.09}_{-0.18}$& $0.04^{+0.18}_{-0.20}$ & $1.77^{+0.08}_{-0.08}$&$207^{+20}_{-9}$&$1215^{+492}_{-569}$& $2.8 \times 10^{10}$\\
     \hline
     {PDS 70 c}&$1032^{+32}_{-21}$& $3.23^{+0.22}_{-0.16}$& $0.09^{+0.12}_{-0.14}$ & $2.35^{+0.19}_{-0.17}$&$-$& $-$& $-$ \\
     {}& $1046^{+42}_{-28}$& $3.26^{+0.28}_{-0.18}$& $-0.04^{+0.14}_{-0.20}$ & $2.22^{+0.20}_{-0.18}$&$230^{+55}_{-62}$&$167^{+146}_{-114}$& 0.88\\
     \hline
     \enddata
     \footnotesize{Notes: The median along with the 16th and 84th percentiles of the marginalized posteriors are reported. B$_{12}$ is the Bayes factor between the blackbody CPD model and the no blackbody model.}
\end{deluxetable*}

\section{The Brighter-Fatter Effect}
\label{sec:bfe}

\begin{figure*}
\includegraphics[width=\linewidth]{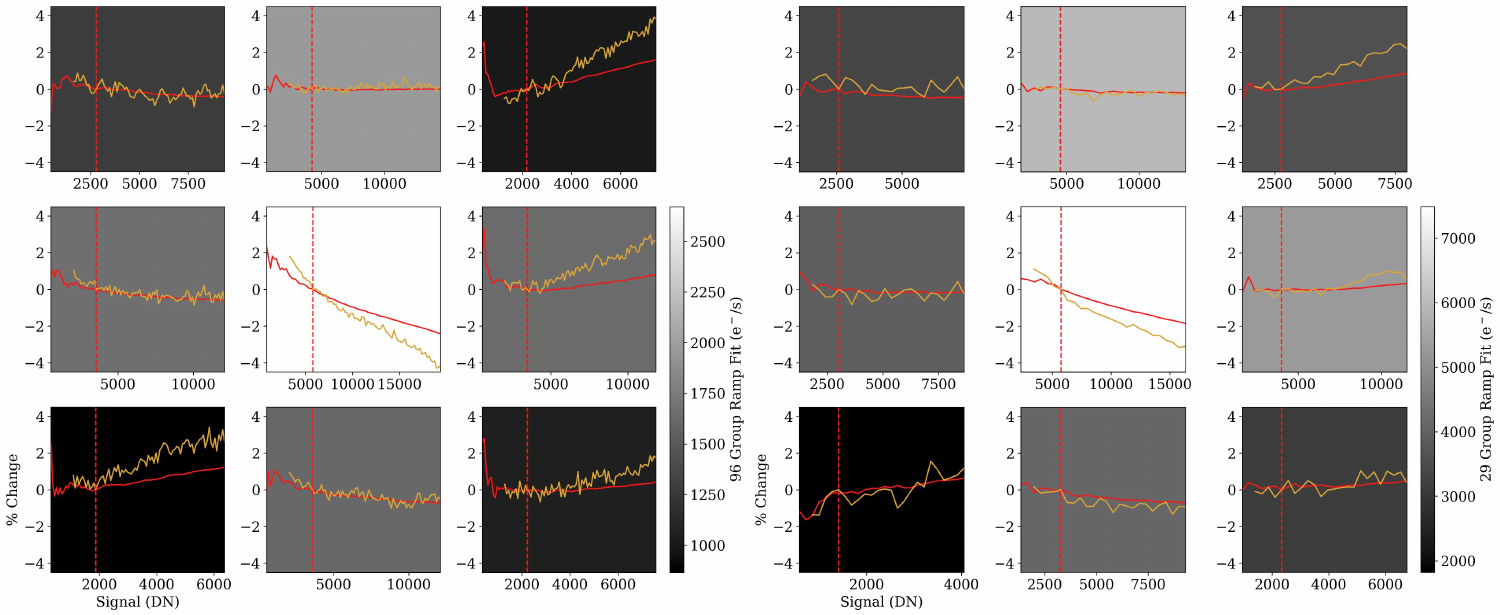}
\caption{Evolution of the central 3$\times$3 pixels as a function of the intensity in each pixel (in the linearized-ramp level data). {Each panel represents a pixel, and is coloured by the count rate in that pixel from the ramp fit to the entire dataset}. PDS 70 is shown on the left and HD 123991 is shown on the right. The solid red line shows the evolution of the mean of the \texttt{calints} frames, after data cleaning. Here we are plotting the calculated rate minus the rate calculated from the first 28 groups and the first 10 groups, for PDS 70 and HD 123991, respectively. These rates are divided by the rate of the ramp fit using all of the groups. The {orange} line shows the evolution of the linearized ramp-level data (ramp-level immediately after it has undergone the \texttt{linearity} step, in stage 1 of the \texttt{jwst} pipeline). To clearly illustrate the change in rate in each pixel as a function of well-depth, we are plotting the difference between groups separated by 15 groups and 5 groups for PDS 70 and HD 123991, respectively, for increasing signal. From this, we additionally subtract the difference between the 25th and 10th group for PDS 70 and the difference between the 10th group and the 5th group for HD 123991 and then normalize this quantity by the difference between the final group and one 15 and 5 groups earlier for PDS 70 and HD 123991, respectively. The vertical dashed red line shows the extent of the data that were used in our analysis.}
\label{fig:pix_ev}
\end{figure*}

In Figure \ref{fig:pix_ev}, we show the evolution of the pixel intensities for the central 9 pixels of the PSF, for PDS 70 and HD 123991. We show how the linearized ramp data evolves {(orange)}, as each pixel accumulates signal, along with how this affects the calibrated rate data (red), output by the \texttt{jwst} pipeline (the \texttt{calints} data). We note that for the central pixel, the detected signal decreases, apparently at all signal levels, with the signal in most of the adjacent pixels either rising or falling, with noticeable differences between PDS 70 and HD 123991. The chosen cutoff value, where we discard all data above this group, is denoted by the vertical dashed red line. For all pixels other than the central pixel, there is no significant signal deficit/excess up to this cutoff value. 



\end{document}